\documentclass[aps,rmp,amsmath,amssymb,twocolumn,graphicx,endfloats,longbibliography]{revtex4-1}
\usepackage{bm}
\usepackage{bbold}
\usepackage{color}
\usepackage[title,toc]{appendix}
\usepackage{xcolor}
\usepackage{natbib}

\newcommand{\ket}[1]{\left|#1\right\rangle}

\begin{document}

\title{Modes and states in quantum optics}

\author{C. Fabre}
\email{claude.fabre@lkb.upmc.fr}
\author{N. Treps}
\email{nicolas.treps@lkb.upmc.fr}

\affiliation{Laboratoire Kastler Brossel, Sorbonne Universite,
ENS, CNRS, College de France, Campus Pierre et Marie Curie, \\ 75005 Paris, France}

\begin{abstract}

A few decades ago, quantum optics stood out as a new domain of physics by exhibiting states of light with no classical equivalent. The first investigations concerned single photons, squeezed states, twin beams and EPR states, that involve only one or two modes of the electromagnetic field. The study of the properties of quantum light then evolved in the direction of more and more complex and rich situations, involving many modes, either spatial, temporal, frequency, or polarization modes. Actually, each mode of the electromagnetic field can be considered as an individual quantum degree of freedom. It is then possible, using the techniques of nonlinear optics, to couple different modes, and thus to build in a controlled way a \textit{quantum network} \cite{Kimble2008} in which the nodes are optical modes,  and that is endowed with a strong multipartite entanglement. In addition, such networks can be easily reconfigurable and subject only to weak decoherence. They open indeed many promising perspectives for optical communications and computation.

Because of the linearity of Maxwell equations a linear superposition of two modes is another mode. This means that a "modal superposition principle" exists hand in hand with the regular quantum state superposition principle. The purpose of the present review is to show the interest of considering these two aspects of multimode quantum light in a global way. Indeed using different sets of modes allows  to consider the same quantum state under different perspectives: a given state can be entangled in one basis, factorized in another.  We will show that there exist some properties that are invariant over a change in the choice of the basis of modes. We will also present the way to find the minimal set of modes that are needed to describe a given multimode quantum state. We will then show how to produce, characterize, tailor and use multimode quantum light, consider the effect of loss and of amplification on such light and the modal aspects of the two-photon coincidences. Switching to applications to quantum technologies, we will show in this review that it is possible to find not only quantum states that are likely to improve parameter estimation, but also the optimal modes in which these states "live". We will finally present how to use such quantum modal networks for measurement-based quantum computation. 

\end{abstract}

\maketitle

\tableofcontents

\section{Introduction}

Modes of the electromagnetic field have always been an important tool, both for theory and applications, helping physicists to describe and use subtle properties of light. For example, optical  telecommunications, in the search of increasing further the number of degrees of freedom used to encode information in a single beam of light, was led to the multiplexing of more and more optical modes, successively temporal modes, frequency modes and transverse modes.

Early research in Quantum Optics concentrated on simple non-classical states of light,  like single photons, squeezed states, twin beams and EPR states, that involved only one or two modes of the electromagnetic field. Then, quantum states involving more and more quantum degrees of freedom were considered. This was implemented either by increasing the size of Fock state basis in a given mode (which is not the subject of the present review), or by increasing the number of modes on which the quantum state extends, these modes being spatial modes for the study the quantum properties of optical images, or time/frequency modes to investigate the quantum properties of light pulses. 

In addition to the intrinsic interest of exploring further the quantum aspects of light in all its manifestations, the study of highly multimode  quantum light is of great interest in the perspective of applications in the domains of quantum technologies. For example, in order to be competitive with present classical computers, the future quantum computer will necessarily make use of quantum states displaying entanglement between many degrees of freedom, and multimode quantum states of light are promising candidates to achieve this goal. 

This review paper aims at characterizing from a very general perspective the numerous specific features and interests of multimode quantum light. Indeed, Quantum Optics, as the child of Optics and Quantum Mechanics, has inherited a double linearity: that of Maxwell equations, for which optical modes form a basis of solutions, and that of Quantum Mechanics, which involves bases of quantum states. One is naturally led to use two "intertwined" Hilbert spaces to describe multimode quantum light: that of modes, and that of quantum states. The properties of single photon quantum states, for example, are intimately related to the characteristics of the mode in which they are defined. 

The specific approach of this paper is to consider these two intricate aspects of quantum light on an equal footing. Note that in a given physical system, there are many possible bases of optical modes: one has the choice of the basis of modes used to describe it. Using different sets of modes allows us to consider the same quantum state under different perspectives: for example a given quantum state can be entangled in one basis, and factorized in another. One can also derive and use the basis of so-called "normal modes", or look for 'principal modes', which contain most of the information of the multimode system.

We will show that there exist some properties that are invariant over a change in the choice of the basis of modes, like non-classicality, non-Gaussianity, negativity of the Wigner function and total photon number. We will also present the way to find the minimal set of modes that are needed to describe a given multimode quantum state, an important tool to reduce the size of the Hilbert space of the system. We will describe how to use modal noise correlation matrices to extract "principal modes" that concentrate and simplify the relevant information over complicated multimode states, especially (but not only) in the case of Gaussian noise. We will show that there always exists a mode basis on which multimode entangled Gaussian states, pure or mixed, are separable.  We will detail the different detection techniques allowing experimentalists to determine such correlation matrices, as well as the ways to generate multimode light and tailor the spatio-temporal shapes of the modes. The effect of loss and of amplification on multimode non-classical light will also be presented. Switching to applications to quantum technologies, we will also show that it is possible to find not only quantum states that are likely to improve parameter estimation, but also the optimal modes in which these states "live". We will finally present how to use such quantum modal networks for measurement-based quantum computation. 

This review will deal with both discrete variables and continuous variables, with an emphasis on the latter, i.e. on the properties of quantum field fluctuations. In addition, we will not treat polarization and angular momentum modes, which have been the subject of intense research \cite{Barnett2016,Korolkova2002}. We will not treat either the multimodal aspects of light matter interaction  \cite{Afzelius2009,Nunn2008}. We will be interested more in "scalable" properties, i.e. features that can be readily extended to large numbers of modes, than in two-mode systems. Even with these restrictions, the domain is vast and the object of numerous theoretical and experimental investigations, so that we will not claim to present in this paper an exhaustive account of it and describe what we consider the most striking features of the domain. There are already a number of review papers and books that cover parts of this domain \cite{Bruss2019,Ou2017,Grynberg2010,Braunstein2012,Weedbrook2012,Adesso2014,Adesso2007,Ansari2018,Leuchs2019,Furusawa2015,Simon1994}, but, to the best of our knowledge, no fully comprehensive presentations of it.

\section{Description of classical multimode light}\label{S1}

In this paper, we will use the complex representation, or analytical signal, ${\bm E}^{(+)}({\bm r}, t)$,  of a classical electric field, such as the real field,  ${\bm E}({\bm r}, t)$, is the vector equal to ${\bm E}^{(+)}({\bm r}, t)+({\bm E}^{(+)}({\bm r}, t))^*$.   

\subsection{Mode of the electromagnetic field and mode basis: definitions}

We will call \textit{mode of the electromagnetic field a vector field ${\bm f}_1({\bm r}, t)$  which is  a normalized solution  of Maxwell equations in vacuum.} It satisfies the three following equations:
\begin{equation}\label{maxwell}
(\Delta -\frac1{c^2}\frac{\partial^2}{\partial t^2}){\bm f}_1({\bm r}, t)= 0, \quad {\bm \nabla} \cdot {\bm f}_1({\bm r}, t)=0 
\end{equation}
and at any time $t$:
\begin{equation} 
\frac1{V}\int_V d^3r \,   |{\bm f}_1({\bm r}, t)|^2=1
\label{mode1}
\end{equation}
$V$ being the very large volume which contains the whole physical system under consideration.

Starting from a given mode, which can have any shape in space and time provided it fulfills equations (\ref{mode1}), one can always build a \textit{orthonormal mode basis} $\{{\bm f}_m({\bm r}, t)\}$ on which one can decompose any solution of the Maxwell equations and which has the function  ${\bm f}_1({\bm r}, t)$ as its first element. As we have restricted ourselves to the case where the field of interest is enclosed in a spatial box of size V, this basis is discrete and can be labeled by a set of integers globally named  as $m$, so that one can write any complex field ${\bm E}^{(+)}({\bm r},t)$ as:
\begin{equation}\label{clas1}
{\bm E}^{(+)}({\bm r},t)=\sum_{m} \mathcal E_{m}{\bm f}_{m}({\bm r},t)
\end{equation}
with, at any time $t$:
\begin{equation}\label{clas2}
\frac1{V}\int_V d^3r \,   {\bm f}_m^*({\bm r}, t)\cdot {\bm f}_{m'}({\bm r}, t)=\delta_{m\, m'}
\end{equation}
$ \mathcal E_{m}$ being the complex amplitudes of the different modes that completely define a given field. It will be often useful to consider the \textit{field quadratures} $E_{mX}$ and $E_{mP}$ as the real and imaginary components of  $ \mathcal E_{m}$:
\begin{equation}\label{quad}
 \mathcal E_{m}=E_{mX}+ i E_{mP}
\end{equation}

\subsection{Hilbert space of modes}

Equation (\ref{clas1}) shows that a given solution of Maxwell equations ${\bm E}^{(+)}({\bm r},t)$ can be considered as a vector belonging to a Hilbert space, called "modal space", that we will denote by an arrow $\overrightarrow{E}$:
\begin{equation}\label{modeclas}
\overrightarrow{E}= \sum_m \mathcal{E}_m \overrightarrow{f_m}
\end{equation}
the sum being in practice limited to a finite number of modes $N$. $\overrightarrow{f_m}$ is a unit vector in the modal space, i.e. a column vector of zeros except for a $1$ at the $m^{th}$ position. $\overrightarrow{E}$ is therefore the column vector $( \mathcal{E}_1,  \mathcal{E}_2, ...)^T$, $T$ denoting the transposition operation in the modal Hilbert space.  
Note that the modal column vector $ \overrightarrow{f_m}$ and the electric field ${\bm f}_m({\bm r}, t)$, function of ${\bm r}$ and $t$, are two representations of the same physical mode. We will thus often denote the mode as ${\bm f}_m$, a notation independant of any representation.

The inner product is defined as usually: 
\begin{equation}\label{clasn2}
\overrightarrow{F}^{T*}  \cdot \overrightarrow{G}= \sum_{m=1}^ N \mathcal{F}^*_m \mathcal{G}_m 
\end{equation}
$\overrightarrow{F}^{T*} = (\overrightarrow{F}^{T})^*$ being the line matrix dual of $\overrightarrow{F}$.  The inner product is also equal to the spatial overlap between the two associated electric fields:
\begin{equation}\label{clasn1}
\overrightarrow{F}^{T*}   \cdot \overrightarrow{G}= \frac1{V}\int_V d^3r \,   {\bm F}^*({\bm r}, t) \cdot {\bm G}({\bm r}, t)
\end{equation}
Note that the value of the inner product (\ref{clasn1}) does not depend on time.

In the following, we will extend the notation with an arrow to other column vectors of dimension $N$, for example $\overrightarrow{\hat{A}}$, of components  $\hat{A}_m$ which are quantum operators. $\overrightarrow{\hat{A}}^T $ denotes the corresponding line vector of operators $\hat{A}_m$, the transposition operation $T$ acting only in the modal space leaving operators unchanged. In the same spirit  $\overrightarrow{\hat{A}}^{\dagger}$ is the column vector of components   $\hat{A}_m^{\dagger}$, the operation denoted by symbol $\dagger$ being used only for the Hilbert space of quantum operators.

\subsection{Different mode bases for the electromagnetic field}

Let us now consider a \textit{modal unitary transformation } $U$ of complex components  $U_n^{m}$, and let us define the modal vectors
\begin{equation}\label{U}
 {\overrightarrow g}_{n}=\sum_{m=1}^N U_n^{m} \, {\overrightarrow f}_{m} 
\end{equation}
It is easy to show that they form another complete set of orthonormal modal vectors (with $\overrightarrow{g}^{T*}_n \cdot \overrightarrow{g}_{n'}=\delta_{n,n'}$), and hence a new basis on which any field can be expanded:
\begin{equation}\label{clas3}
{\bm E}^{(+)}({\bm r},t)=\sum_{n} \mathcal G_{n}{\bm g}_{n}({\bm r},t) \quad ; \quad \overrightarrow{E}=\sum_{n} \mathcal G_{n} \overrightarrow{g}_n
\end{equation}
The introduction of modes made in this section has been so far purely mathematical. It leaves us a total freedom of choice of the unitary transformation $U$, and therefore of the mode basis. The most widely used is the basis of plane wave modes, that is easy to handle mathematically, but rather unphysical. There are other mode bases that fit better the light source and the optical system under consideration. Let us quote some of them:
\begin{itemize}

\item spatial Hermite-Gauss modes are well suited for beams produced in cavities made of spherical mirrors. Laguerre-Gauss modes, carrying orbital angular momentum, are used in quantum information processing, in particular in Quantum Key Distribution context \cite{Mirhosseini2015};

\item temporal or frequency Hermite-Gauss modes provide a convenient orthogonal basis for light pulses of any temporal and spectral shape \cite{Brecht2015};

\item tilted plane waves are used to describe "images" i.e. any field configuration in the transverse plane, perpendicular to the main propagation direction \cite{Kolobov2006}.

\end{itemize}

 The choice of the mode basis can also be adapted to the detection process:
 \begin{itemize}
 
 \item the \textit{pixel mode basis} is useful when one considers imaging devices such as a CCD camera or a photodiode array. It consists of spatial modes which are of constant value over the pixel area $\delta x \times \delta x$ and zero outside. Such a basis is orthonormal but not complete; 
 
 \item the  \textit{time bin basis} is analogous to the previous one when one replaces space by time and is useful to analyze temporal sequences;
 
 \item in the same spirit, the  \textit{frequency band basis} consists of frequency bins of width $\delta \nu$, useful to analyze broadband sources.
 
 \item the \textit{sideband mode basis}, consisting of sums of monochromatic frequency modes symmetrically disposed around a carrier frequency, is useful to describe the different Fourier components of a time-dependent signal (see section X C).
   
 \end{itemize}   
 
 \subsection{Transverse and longitudinal modes}\label{paraxial3}
 
For the sake of simplicity, we will now restrict our analysis to the case where the fields of interest are the superposition of plane waves having wavevectors close to a mean value ${\bm k}_0$ (paraxial approximation) and frequencies close to a central frequency $\omega_0= c |{\bm k}_0|$ (narrowband approximation). To simplify notations, we assume that ${\bm k}_0$ is parallel to the $z$ axis. The electric field can be written as:
\begin{equation}\label{paraxial1}
{\bm E}^{(+)}({\bm r},t)=  e^{i(k_0 z-\omega_0 t)} \sum_{m} \mathcal{F}_{m}{\bm f}_{m}({\bm r},t)
\end{equation}
$e^{i(k_0 z-\omega_0 t) }$ is the carrier plane wave and ${\bm f}_{m}({\bm r},t)$ the envelope functions of the different modes, which are slowly varying functions of time at the scale of the optical period and of position at the scale of the wavelength.

Let us now simplify further our approach by restricting the set of unitary modal transformations $U$ to the ones that are factorized in transverse and longitudinal factors, the polarization degree of freedom being unaffected. This allows us to use factorized modes ${\bm f}_ {m}$:
\begin{equation}\label{paraxial2}
{\bm f}_ {m}({\bm r},t)={\bm \epsilon}_{i}  f^{(T)}_p(x, y, z) f^{(L)}_r (t, z)
\end{equation} 
$ {\bm \epsilon}_{i} $ ($i=1,2$) are two orthogonal polarization unit vectors in the $xy$ plane,  $f^{(T)}_p$ is  the transverse (or spatial) part of the mode, and $f^{(L)}_r$ its longitudinal (or temporal) part, $m$ being a short notation for $(i,p,r)$.

To build a spatial mode basis, one can use any orthonormal set of functions of  $x,y$,  $f_p(x, y) $, provided they vary slowly on the wavelength scale. The transverse mode functions $f^{(T)}_p (x, y, z) $  for any value of $z$ are then deduced from their values at $z=0$ $f^{(T)}_p (x, y, z=0)=f_p(x, y) $ by a Fresnel integral accounting for diffraction.

To build a temporal mode basis, one can use any set  of functions of time $t$, $f_r (t)$, provided they satisfy the slowly varying envelope approximation and temporal orthonormality relations with an integration time $T$ longer than any other characteristic time of the problem. The longitudinal mode functions are then $f^{(L)}_r (t,z)=f_r(\tau=t-z/c)$, describing undistorted  pulses in the reference frame propagating at the velocity of light in the $z$ direction.

The three factors in Equation (\ref{paraxial2}) correspond to three different kinds of quantum multimode effects: polarization effects, spatial effects and temporal effects. They appear here as decoupled, because of our simplifying assumptions. We will not detail the quantum properties linked to the polarization of light, and in particular the modes carrying orbital and polarization angular momentum.  They are treated in detail in \cite{Korolkova2002,Barnett2016}.  In this review we will consider a single polarization component, drop the polarization index $i$ and work with scalar modes. Such a description in terms of decoupled spatial and temporal modes is convenient when one treats problems where the temporal shape of light does not modify its transverse properties, or when diffraction does not modify the pulse shape. Self-focussing effects, either in space or time, or objects like X-waves \cite{Gatti2009,Jedrkiewicz2012} would clearly need a more elaborate approach.

When a single temporal mode $f^{(L)}_1(t, z)$ is involved, the electric field writes: 
\begin{equation}\label{image}
{\bm E}^{(+)}({\bm r},t)=   {\bm \epsilon}_{1} e^{i(k_0 z-\omega_0 t) } f^{(L)}_1(\tau) \sum_{p} {\mathcal E} _{p} f^{(T)}_{p}({\bm r})  
\end{equation}
Omitting the factor in front of the sum, one defines the transverse electric field as:
\begin{equation}\label{image2}
E^{(+)}_{T}(x,y,z)=  \sum_{p}  {\mathcal E}_{p}  f^{(T)}_{p}(x,y,z)
\end{equation}
Its Fourier transform in transverse wavevector space, ${\tilde E}^{(+)}_{T}(k_x, k_y)$, can be also expanded on the basis of  $k_x$, $k_y$ dependent modes, $\tilde f^{(T)}_{p}(k_x, k_y)$:
\begin{equation}\label{image3}
{\tilde E}^{(+)}_{T} (k_x, k_y)= \sum_p   {\tilde {\mathcal E}}_{p} \tilde f^{(T)}_{p}(k_x, k_y)
\end{equation}
It is independent of $z$ as diffraction does not modify the distribution of transverse wave-vectors. Transverse modes are well suited to the description of {\it quantum imaging} problems \cite{Kolobov2006}, i.e. the quantum properties of the transverse distribution of light, for example quantum correlations between different points of the transverse plane.

In a symmetrical way, when a single spatial mode $f^{(T)}_1({\bm r})$ is involved, the electric field writes: 
\begin{equation}\label{pulse}
{\bm E}^{(+)}({\bm r},t)=  {\bm \epsilon}_{1} e^{-i \omega_0 \tau } f^{(T)}_1({\bm r}) \sum_{r} {\mathcal E} _{r} f^{(L)}_{r}(\tau)
\end{equation}
This allows us to define the temporal or longitudinal field as:
\begin{equation}\label{pulse2}
E^{(+)}_{L}(t,z) =  \sum_{r} {\mathcal E} _{r}  f^{(L)}_{r}(\tau) 
\end{equation}
Its Fourier transform ${\tilde E}^{(+)}_{L}(\omega)$ can be expanded on a basis of frequency modes $\tilde f^{(L)}_{r}(\omega)$:
\begin{equation}\label{pulse3}
{\tilde E}^{(+)}_{L}(\omega) =  \sum_{r} {\tilde {\mathcal E}} _{r}  {\tilde f}^{(L)}_{r}(\omega) 
\end{equation}
Temporal/frequency modes are well suited to the quantum description of light pulses and of their correlations. \textit{Frequency combs} are an important particular case. They can be expanded on the basis of monochromatic waves of equally spaced frequencies or, in the time domain, as trains of identical pulses equally spaced in time.

 \subsection{Modes and classical coherence}
 
Note that if the classical field under study is totally defined, i.e. is perfectly coherent, in both time and s  pace, its normalized spatio-temporal shape can be taken as a first mode of a mode basis, and the decomposition (\ref{clas1}) comprises a single term: \textit{any perfectly coherent classical field is single mode in essence}. This is the case for example for a mode-locked laser in which the relative phases between the different frequency modes are fixed.

But in most practical cases, the field is not perfectly mastered: it has some degree of randomness or "incoherence" in the form of amplitude and phase  fluctuations at any point of space and time. In this situation, the complex coefficients $\mathcal{E}_m$ in (\ref{clas1}) are \textit{stochastic quantities}. This is the case for example for a multimode c.w. laser in which the phases of each individual frequency component are randomly fluctuating. The classical fluctuations of the field have no reason to be described by a single fluctuating amplitude $\mathcal{E}_1$, so that a full decomposition of the form  (\ref{clas1}) is indeed needed, and the degree of coherence of the field is characterized by the probability distributions of the complex amplitudes $\mathcal{E}_m$ and by the correlations existing between different amplitudes $\mathcal{E}_m$  and $\mathcal{E}_{m'}$. 

Among the different quantities that are used to characterize the degree of coherence of a classical stochastic field, we will use the following matrices characterizing its fluctuations:
\begin{itemize}

\item 1) The \textit{first order coherency matrix} ${\bf {\Gamma}}^{(1)}$ \cite{Wiener1927}, defined by:
\begin{equation}\label{coherence}
\left({\bf {\Gamma}}^{(1)} \right)_{m,n}=\overline{{\mathcal{E}_m^* \mathcal{E}_n}} \quad ; \quad {\bf \Gamma}^{(1)} = \overline{\overrightarrow E \overrightarrow{E}^{T*} },
\end{equation}
the bar indicating an ensemble average. It allows us to determine the cross-correlation function for the field amplitudes \cite{Barakat1963}:
\begin{eqnarray}\label{G1}
G^{(1)}({\bm r}, {\bm r'},t,t') &=&\overline{E^{(+)*}({\bm r'},t') E^{(+)}({\bm r},t)}\\
&=&\sum_{m,n} \left({\bf {\Gamma}}^{(1)} \right)_{m,n}  f_m^*({\bm r'},t') f_n ({\bm r},t)
\end{eqnarray}
 The coherency matrix has been extensively studied in the context of polarization modes \cite{Refregier2005} and imaging \cite{Yamazoe2012}.

\item 2) The \textit{quadrature covariance matrix} ${\bf \Gamma}_Q$, :

Let us define
\begin{eqnarray}\label{q1}
\overrightarrow Q&=& \nonumber\\
& & (E_{1X}, E_{2X}, \ldots, E_{NX},  E_{1P},  E_{2P}, \ldots, E_{NP})^T
\end{eqnarray}
the column vector containing all field quadratures.  The quadrature covariance matrix ${\bf \Gamma}_Q$, that we will call briefly \textit{"covariance matrix"}, is the real  $2N \times 2N$ matrix defined on a given mode basis $\{\overrightarrow f_n\}$ as:
\begin{equation}\label{compcov}
{\bf \Gamma}_Q = \overline{\overrightarrow Q \overrightarrow{Q}^{T} }
\end{equation}
It contains all the second moments of the different modes, $\overline{E_{mX}^2}$  and $\overline{E_{mP}^2}$ on the diagonal, and outside the diagonal all the quadrature correlations $\overline{E_{mX} E_{nX}}$, $\overline{E_{mP} E_{nP}}$
and $\overline{E_{mX} E_{nP}}$.

The quadrature covariance matrix is a symmetric real matrix of size $2N \times 2N$, where $N$ is the number of modes, whereas the coherency matrix is hermitian of size $N \times N$. If all the modes have fluctuations with Gaussian statistics and zero mean, then \textit{the quadrature coherence matrix contains all  the physical information about the system}, which is not the case of the first order coherency matrix. 

The $N \times N$ coherency matrix can be deduced from the $2N \times 2N$ quadrature covariance matrix, using the relations:
\begin{eqnarray}\label{compcovx} 
({\bf \Gamma}^{(1)})_{m,n} &=& \overline{E_{mX} E_{nX}} + \overline{E_{mP} E_{nP}}\nonumber \\
&+&i(\overline{E_{mX} E_{nP}}  -\overline{E_{mP} E_{nX}})
\end{eqnarray}

In contrast, one cannot determine the quadrature covariance matrix from the  coherency matrix. This last matrix gives information about the distribution of energy among the different modes, but not the way it is distributed between the two quadratures inside a given mode.

\end{itemize}

\section{Description of quantum multimode light}

Let us now consider quantum fields, and start with the usual approach of Quantum Electrodynamics \cite{Cohen1987,Mandel1995,Grynberg2010}  which consists in introducing the electric field operator in the Heisenberg representation ${\hat{\bm E}}^{(+)}({\bm r},t)$ as the quantum extension of the classical complex field ${\bm E}^{(+)}({\bm r},t)$ and to expand it on the basis of \textit{monochromatic plane wave modes} ${\bm u}_{\ell}({\bm r},t)$:
\begin{eqnarray}\label{dec}
{\hat{\bm E}}^{(+)}({\bm r},t)&=&\sum_{\ell}\mathcal E_{\ell}^{(1)} \, \hat{a}_{\ell} {\bm u}_{\ell}({\bm r},t), \nonumber \\
{\bm u}_{\ell}({\bm r},t)={\bm \epsilon}_{\ell} e^{i ({\bm k}_{\ell} \cdot {\bm r}-\omega_{\ell} t)} && \mathcal E_{\ell}^{(1)} =\sqrt{\frac{\hbar \omega_{\ell}}{2 \varepsilon_0 V}}
\end{eqnarray}
where ${\bm \epsilon}_{\ell}$ is a unit polarization vector, $\hat{a}_{\ell} $ the annihilation operator of a photon in the plane wave mode and $ \mathcal E_{\ell}^{(1)}$ the single photon electric field. The set of plane wave modes $\{{\bm u}_{\ell}({\bm r},t)\}$ satisfies the orthonormality condition (\ref{clas2}) and the equal time completeness relation in space domain, valid at any time $t$:
\begin{equation}\label{close}
 \sum_{\ell} {\bm u}^*_{\ell}({\bm r},t) \cdot {\bm u}_{\ell}({\bm r}',t)= 2V \delta^{(3)}({\bm r}-{\bm r}')
\end{equation}
where $\delta^{(3)}$ is the delta function in 3-dimensional space.

Expansion (\ref{dec}) shows that the modal Hilbert space of classical electromagnetic fields can be mapped into a modal Hilbert space of quantum field operators, so that one can write:
\begin{equation}\label{qu1}
\overrightarrow{ {\hat E}}=\sum_{\ell}  \mathcal E_{\ell}^{(1)}  \hat{a}_{\ell} \overrightarrow{u_{\ell}}
\end{equation}
$\overrightarrow{ {\hat E}}$ is thus the column vector of operators $\mathcal E_{\ell}^{(1)}  \hat{a}_{\ell}$.
Note that the electric field quantum operator ${\hat{\bm E}}^{(+)}({\bm r},t)$ in the Heisenberg representation obeys Maxwell equations (\ref{maxwell}), which form the basis, not only of classical electrodynamics, but also of quantum electrodynamics.

\subsection{Electric field operator in any mode basis}

Let us now perform a modal unitary transformation $U$ on the set of creation operators $\{\hat{a}^{\dagger}_{\ell}\}$, yielding a new set of operators $\{\hat{b}^{\dagger}_{m}\}$ given by:
\begin{equation}\label{U1}
{\hat b}^{\dagger}_{m}=\sum_{\ell}U_m^{\ell} \, \hat{a}^{\dagger}_{\ell} \quad ; \quad  \overrightarrow{ {\hat b}^{\dagger}}= U \overrightarrow{ {\hat a}^{\dagger} }
\end{equation}
where $\overrightarrow{ {\hat b}^{\dagger}}$ and $\overrightarrow{ {\hat a}^{\dagger}}$ are column vectors of components ${\hat b}^{\dagger}_{m}$ and ${\hat a}^{\dagger}_{\ell}$. One has also:
\begin{equation}\label{U2}
\hat{a}_{\ell}=\sum_m U_m^{\ell} \, \hat{b}_m \quad ; \quad  {\overrightarrow {\hat a}}= U ^T{\overrightarrow {\hat b}} 
\end{equation}
Note that the mode $\overrightarrow{u_l}$, in its column vector representation, is associated with the single creation operator ${\hat a_l}^{\dagger}$ while the mode basis $\{\overrightarrow{u_l}\}$ is associated with the column vector of creation operators $\overrightarrow{ {\hat a}^{\dagger}}$.

The unitarity of matrix $U$ ensures that:
\begin{equation}
 [\hat{b}_{m}, \hat{b}^{\dagger}_{m'}]= \delta_{m,m'} \quad or \quad \overrightarrow{ {\hat b}}(\overrightarrow{\hat b^\dagger})^T - \overrightarrow{ \hat b^\dagger} (\overrightarrow{\hat b})^T =  \hat{1}
\end{equation}
The operators $\hat{b}_{m}$ are indeed bosonic operators, and the positive electric field operator can now be written as a linear combination of the annihilation operators $\hat{b}_{m}$ in a way similar to the decomposition (\ref{dec}) or  (\ref{qu1}):
\begin{equation}\label{decomp}
{\hat {\bm E}}^{(+)}({\bm r},t)=\sum_{m} \mathcal F_m^{(1)} \hat{b}_{m} {\bm f}_{m}({\bm r},t) \, ; \, 
\overrightarrow{ {\hat E}}=\sum_{m}  \mathcal F_{m} ^{(1)}\hat{b}_{m} \overrightarrow{f_{m}}
\end{equation}
${\hat b}_{m}$ is the annihilation operator of one photon in the normalized mode ${\bm f}_{m}({\bm r}, t)$ (that we will write more simply as ${\bm f}_{m}$), and the electric field per photon $\mathcal F_m^{(1)}$ is given by:
\begin{equation}
(\mathcal F_m^{(1)})^2 = \sum_{\ell} (\mathcal E_{\ell}^{(1)})^2 \,  | U_{m}^{\ell}|^2 
\end{equation}
The column vector $\{ \overrightarrow{f_{m}}\}$ contains a new set of modes on which the field is expanded. It is related to the plane wave basis by:
\begin{equation}\label{sum}
{\overrightarrow f}_m=\frac1{\mathcal F_m^{(1)} }\sum_{\ell} \mathcal E_{\ell}^{(1)}  U_m^{\ell}{\overrightarrow u}_{\ell}
\end{equation}

We have therefore shown how to write in the most general case the quantum field on any mode set $\{ \overrightarrow{f_{m}}\}$ and to define the associated annihilation operators $\hat{b}_{m}$. Strictly speaking, because of the presence of the frequency dependent scaling factor $\mathcal E_{\ell}^{(1)}$ in the sum (\ref{sum}), these new modes are not  necessarily orthogonal when their frequency spectrum is very broad. They are indeed orthogonal when the unitary modal transformation ${\bf U}$ mixes only plane waves oscillating at nearby frequencies $\omega \simeq \omega_0$, (narrowband approximation studied in section II.D), in which case ${\mathcal F_m^{(1)}}={\mathcal E_{\ell}^{(1)}}(\omega_0)$, so that one can very simply write:
\begin{equation}\label{sum2}
{ \overrightarrow f}_m=\, \sum_{\ell} U_m^{\ell}{\ \overrightarrow u}_{\ell} 
\end{equation}
We note that in this particular case, the transformation (\ref{sum2}) for the mode shape and  the transformation (\ref{U}) for the creation operators are identical. Using (\ref{sum2}), one easily shows that the completeness relation (\ref{close}) holds also in this case in the new mode basis. 

In a way related to equation (\ref{quad}), one can write the field operator (\ref{decomp}) in terms of hermitian \textit{dimensionless quadratures operators} ${\hat X}_m$ and ${\hat P}_{m}$ in the different modes  ${\bm f}_m$,  such that:
\begin{eqnarray}\label{quadquant}
\hat{b}_m &=&(\hat{X}_m + i \hat{P}_m)/2 \nonumber\\
 \overrightarrow{ {\hat E}}&=&\sum_m \mathcal{F}^{(1)}_m \frac{{\hat X}_m + i {\hat P}_m}{2} { \overrightarrow f}_m, \nonumber \\ \hat{X}_m= \hat{b}^{\dagger}_m + \hat{b}_m  &,& \hat{P}_m=i(  \hat{b}_m^{\dagger}- \hat{b}_m )
\end{eqnarray} 
with $[\hat{X}_m,  \hat{P}_m]=2i$. The normalization has been chosen in such a way that the variance of vacuum fluctuations on any quadrature $\hat{X}_m$ or $\hat{P}_m$ is 1. We will also use quadrature operators in a rotated phase space $\hat{X}_{m \phi}$ defined by
\begin{equation}
\hat{X}_{m \phi}= \hat{b}^{\dagger}_{m}  e^{i \phi}  +\hat{b}_{m} e^{-i \phi}=\hat{X}_m\cos \phi +\hat{P}_m \cos \phi 
\end{equation}

 \subsection{The two sides of quantum optics}
 
Let us consider again the quantum form of the modal decomposition of the electric field operator:
\begin{equation}\label{quant}
\overrightarrow{ {\hat E}} = \sum_{m}  \mathcal F^{(1)}_{m} \hat{b}_{m} \overrightarrow{f_{m}}
\end{equation}
It comprises the same modal functions $ \overrightarrow{f_{m}}$ as the classical decomposition (\ref{modeclas}), while the classical complex amplitudes have been replaced by quantum operators. Expression (\ref{quant}) thus exemplifies the \textit{intricate dual nature of light}: the operatorial part $\hat{b}_{m}$ relates it to the Hilbert space of quantum states, its Fock states basis and their interpretation in terms of particles, while the modal part $  \overrightarrow{f_{m}}$ relates it to classical optics, to the modal Hilbert space of solutions of Maxwell equations and to the wave aspect of light. These two aspects are intimately mixed \cite{Xiao2017}, as the annihilation operator $\hat{b}_{m}$ is defined in the specific mode $ {\bm f}_{m}$, which provides the shape in time and space of the probability to detect the photons. 

A striking example of this intimate relation between modes and operators is provided by the following formula,  deduced from (\ref{sum2}), that relates the commutator of annihilation and creation operators ${\hat b}_f$ and ${\hat b}_g^{\dagger}$ associated with any two modes ${\bm f}$ and  ${\bm g}$, even non-orthogonal, to the overlap integral of these modes:
 \begin{equation}\label{comut}
[{\hat b}_f, {\hat b}_g^{\dagger}]= \frac1V \int d^3 r {\bm f}({\bm r}, t)\cdot {\bm g}^*({\bm r}, t)=\overrightarrow{f}^{T*} \cdot \overrightarrow{g}
\end{equation}

These considerations allow us to point out a very important and unique feature of Quantum Optics, its \textit{double linearity}, that of Maxwell equations and that of Quantum Mechanics. It allows us, thanks to the possibility to change mode bases, to consider the same quantum state from different perspectives. This is not the case for other multimode systems, like sets of material qubits: linear combinations of modes are other modes, whereas linear combinations of qubits carried by different systems are not simple physical objects.

It is well-known that coherence is a fundamental notion for the physical domains involving waves: this is of course the case for classical optics \cite{Goodman2015}, but also for quantum mechanics. Coherence of matter waves, and more generally coherence in quantum physics has even been recently considered as a basic resource for quantum technologies \cite{Streltsov2017}. Both coherences must not be confounded and must be indeed taken into consideration in a global way in the domain of quantum optics \cite{Glauber1963,Mandel1995}. They both play important and intricate roles in multimode quantum optics, as we will see more extensively below.

\subsection{Single photon states} 

The bosonic operators  $\hat{b}_m$ allow us to define number operators $\hat{N}_m= \hat{b}_m^{\dagger} \hat{b}_m$ and their eigenstates $|n_m: {\bm f}_m\rangle $, where $n_m$ is an integer, which are the number states  in modes ${\bm f}_m({\bm r}, t)$.
 
 Let us call $|0\rangle$ the vacuum state in the plane wave basis $\{{\bm u}_{\ell}\}$, defined by $\hat{a}_{\ell} |0\rangle=0 \, \forall \ell$. One has also from (\ref{U}) and for all $m$:
 \begin{equation}
  \hat{b}_{m} |0\rangle=\sum_{\ell}  U^{ \ell *}_{m} \hat{a}_{\ell} |0\rangle=0
 \end{equation}
The same state $|0\rangle$ is also the vacuum for the new mode basis $\{{\bm f}_m\}$. This allows us to define the quantum state $|1: {\bm f}_m\rangle$ of a single-photon state in any mode ${\bm f}_m$, and to express it in terms of plane wave single photons as
\begin{equation}\label{photon}
|1: {\bm f}_m\rangle= \hat{b}_m^{\dagger} |0\rangle=\sum_{\ell} U^{ \ell}_{m}  |1: {\bm u}_{\ell} \rangle
\end{equation}
Note that  the same unitary transformation $U$ is used for the creation operators (\ref{U}), for the mode shape (\ref{sum2}) and for the single-photon state (\ref{photon}). 

It is important to stress that a single-photon state $|1: {\bm f}_m \rangle$  does not describe a physical object which exactly looks like a classical particle, because \textit{its properties depend on the mode in which it is defined}. Photons are not simply  "very small bodies emitted from shining substances" \cite{Newton1704}; they must rather be considered as the first excitation of mode ${\bm f}_m$ \cite{Lamb1995}. If the unitary transform $U$ mixes modes of different frequencies, then the single-photon state is no longer an eigenstate of the hamiltonian of energy $\hbar \omega$: it describes a non-stationary "single photon wave-packet" \cite{Titulaer1966}; if the unitary transform $U$ mixes modes of different wavevectors, the single photon state is no longer an eigenstate of the momentum with eigenvalue $\hbar {\bf k}$: it describes a more complex single-photon waveform, for example the dipole mode in which a single photon is spontaneously emitted by an excited atom \cite{Cohen1987}. 

It is easy to show that one has also, for any two single-photon states associated with any two modes $ \overrightarrow{f}$ and  $ \overrightarrow{g}$:
\begin{equation}\label{homo}
\langle 1: {\bm f} | 1: {\bm g} \rangle=\frac1{V}\int_V d^3r \,   {\bm f}^*({\bm r}, t)\cdot {\bm g}({\bm r}, t)= \overrightarrow{f}^{T*} \cdot \overrightarrow{g}
\end{equation}
When dealing with single-photon states, the quantum inner product is equal to the modal inner product. There is therefore an exact mapping between a single-photon quantum state $ | 1: {\bm f} \rangle$ and the corresponding spatio-temporal mode amplitude ${\bm f}({\bm r}, t)$, so that it is often convenient to consider ${\bm f}$ as the "wave function" of the single photon \cite{Smith2007}. 

It is also easy to derive from  (\ref{comut}) the following useful relation:
\begin{equation}\label{annihil}
{\hat b}_{{\bm f}} | 1: {\bm g} \rangle=( \overrightarrow{f}^{T*} \cdot \overrightarrow{g} ) |0 \rangle
\end{equation}
where ${\hat b}_{{\bm f}}$ is the annihilation operator in mode ${\bm f}$.

\subsection{Multimode quantum states}

We can write the most general quantum state of light $|\Psi\rangle$ in a given mode basis as:
\begin{equation} \label{state}
|\Psi\rangle=\sum_{n_1}...\sum_{n_m}... C_{n_1,...,n_m,...}|n_1: {\bm f}_1\rangle \otimes ...\otimes |n_m: {\bm f}_m\rangle \otimes ...
\end{equation}
Knowing that $|n_m: {\bm f}_m\rangle=(\sum_{\ell}U^{\ell}_m \hat{a}_{\ell}^{\dagger} )^{n_m} |0\rangle/\sqrt{n_m!}$ and relations (\ref{U}) and (\ref{U2}) it is straightforward to write $|\Psi\rangle$ in terms of number states in any mode basis of our choice. This implies that, in order to characterize a given multimode quantum state, we have a new degree of freedom to play with, namely the \textit {choice of mode basis}, in addition to the choice of the quantum state basis. 

Let us take as an example a quantum state spanning on two modes ${\bm f}_{1}$ and ${\bm f}_{2}$. Another possible mode basis consists of the symmetric and antisymmetric combinations  ${\bm f}_{\pm}=({\bm f}_{1} \pm {\bm f}_{2})/\sqrt2$. The quantum state  
\begin{equation}\label{2photon}
|\psi \rangle=|1: {\bm f}_{1}\rangle \otimes| 1 : {\bm f}_{2}\rangle
\end{equation}
written in the first basis can, can also be written in the second basis, using relation (\ref{U}), as 
\begin{equation}
|\psi \rangle=(|2: {\bm f}_{+}\rangle \otimes| 0 : {\bm f}_{-}\rangle -|0: {\bm f}_{+}\rangle \otimes| 2 : {\bm f}_{-}\rangle)\sqrt2
\end{equation}
$|\psi \rangle$ is therefore factorized in the first mode basis and entangled in the second. The same property holds for the two-mode continuous variable quantum state $|\psi' \rangle$ which consists of a product of two equally squeezed vacuum states on basis (${\bm f}_{1}, {\bm f}_{2}$). On basis   (${\bm f}_{+}, {\bm f}_-$) it is an "EPR entangled state", like the one studied by Einstein, Podolsky and Rosen in their famous paper \cite{Einstein1935}. In the present context \textit{the fact that a quantum state is entangled or not depends on the choice of the mode basis} \cite{Thirring2011}. This is due to the fact that the physical  system we are interested in has not a unique physical bipartition into an "Alice" part and a "Bob" part, all possible combinations of modes being treated on an equal footing. One can say that $|\psi \rangle$, or $|\psi' \rangle$, describes an intrinsic quantum resource, which manifests itself as a product of non-classical states in one basis, and as entanglement on another.

\subsection{Quantum correlation matrices}

The information about multimodal correlations \cite{Giorgi2011}, classical as well as quantum, is contained in different matrices. We introduce here the ones that are the quantum extensions of the classical matrices defined in the previous section  \cite{Leuchs2002,Takase2019}:
\begin{itemize}

\item 1) The \textit{quantum coherency matrix} ${\bf {\Gamma}}^{(1)}$ \cite{Wiener1927}, defined as the extension of the classical one (Equation (\ref{coherence})), has matrix elements in a given mode basis equal to:
\begin{equation}\label{qcoherence}
( {\bf \Gamma}^{(1)} )_{m,n}=\langle {\hat a}_m^{\dagger}  {\hat a }_n \rangle 
\end{equation}
It can be written in a condensed way as ${\bf \Gamma}^{(1)}=\langle \overrightarrow {\hat a^\dagger} \overrightarrow{\hat a}^T\rangle$. 

${\bf \Gamma}^{(1)}$ is an Hermitian, positive matrix of the Gram type which can be related to the first order coherence properties of the field \cite{Glauber1963,Mandel1995,Refregier2005}. Its trace gives the mean total number of photons in the state.

\item 2) The \textit{quantum covariance matrix} ${\bf \Gamma}_Q$ :

Let us name, in a same way as in (\ref{q1}):
\begin{equation}\label{Q}
\overrightarrow{\hat Q} = (\hat X_1, \hat X_2, \ldots, \hat X_N, \hat P_1, \hat P_2, \ldots, \hat P_N)^T
\end{equation}
the column vector containing all quadrature operators.  The quadrature covariance matrix ${\bf \Gamma}_Q$ \cite{Simon1994} is the real  $2N \times 2N$ matrix defined on a given mode basis ($\overrightarrow f_n$) as:
\begin{equation}\label{compcov1}
{\bf \Gamma}_Q = \frac{1}{2} < \overrightarrow{\hat Q} \overrightarrow{\hat Q}^T + (\overrightarrow{\hat Q}\overrightarrow{\hat Q}^T )^T >
\end{equation}

It contains all the second moments of the quadrature operators $\langle \hat{X}_{m}  \hat{X}_{n} \rangle$, $\langle  \hat{P}_{m}  \hat{P}_{n}\rangle$, $\langle  \hat{X}_{m}  \hat{P}_{n} \rangle$ and $(\langle  \hat{P}_{m}  \hat{X}_{n} \rangle+ \langle   \hat{X}_{n} \hat{P}_{m}\rangle)/2$. It allows to write in a compact way the multimode version of the Heisenberg inequality \cite{Simon1994}:
\begin{equation}\label{ineq}
 {\bf \Gamma}_Q + i \beta =  {\bf \Gamma}_Q +
\left( \begin{array}{cc}
\mathbb{0} &i \mathbb{1} \\
-i\mathbb{1} & \mathbb{0}
\end{array}\right) >0
\end{equation}
where $\beta$ is the symplectic form. This relation is invariant under any symplectic transformation, and in particular under any mode basis change.

Note that the vector $\overrightarrow Q$ is of dimension $2N$, whereas the modal vectors $\vec E$ and $\overrightarrow {\hat a}$ defined in sections $II.B$ and $III.A$ are of dimension $N$.  We will mention the difference when necessary in the following.

The coherency matrix is related to the matrix elements of the quadrature covariance matrix by the following expression, quantum extension of equation (\ref{compcovx}):
\begin{eqnarray}\label{compcov3} 
({\bf \Gamma}^{(1)})_{m,n} &=&\frac14[ \langle \hat{X}_{m}  \hat{X}_{n} \rangle + \langle  \hat{P}_{m}  \hat{P}_{n} \rangle \nonumber \\
&+&i(\langle  \hat{X}_{m}  \hat{P}_{n} \rangle - \langle  \hat{P}_{m}  \hat{X}_{n} \rangle)]
\end{eqnarray} 
In particular:
\begin{equation} 
({\bf \Gamma}^{(1)})_{n,n} = \frac14[\langle \hat{X}_{n}^2 \rangle + \langle  \hat{P}_{n}^2 \rangle -2]
\end{equation} 

\end{itemize}

\subsection{Passive and active mode basis change}

The unitary transformation $U$ that connects one mode basis into another one can be seen in two different ways, one "passive", one "active":

\begin{enumerate}

\item it can be considered as a change of point of view on a given quantum state, in the same way as one can use rotations to see a physical object from different perspectives. This is what we have done so far;

\item it can also be considered as a quantum process implemented by a real physical device and modifying the quantum state of the system by a unitary transformation, in the same way as one can rotate the physical object under consideration and keep a fixed point of view.  $U$ corresponds then to an evolution induced by a hamiltonian which is a linear combination of operators of the form ${\hat a}_i {\hat a}_j^{\dagger}$. It can be shown that this transformation can be implemented by a generalized interferometer that mixes all the input modes by appropriate beamsplitters \cite{Reck1994}.

 \end{enumerate}
 
 We will use these two aspects of mode changes in the following.

\subsection{Intrinsic properties of multimode light states}

Note that, whereas entanglement and factorization of state $|\psi \rangle$ (Eq. \ref{2photon}) depend on the mode basis, the total mean photon number in this state is equal to 2 in the two mode bases. There are therefore properties which do not depend on a special choice of mode basis, and that we will call  \textit{intrinsic}. We expect them to have a stronger physical meaning than the properties which depend on this choice \cite{Refregier2005}.

Let us mention here some physical properties that are intrinsic:

- \textit{The vacuum state} $|0\rangle $ is the same in any mode basis, as we have seen in section IIIc. 

- The operator \textit{"total number of photons"} is defined  by:
\begin{equation}
\hat{N}_{tot}=\sum_{m} \hat{b}^{\dagger}_{m}  \hat{b}_{m} =  \overrightarrow{\hat b^\dagger}^T \cdot \overrightarrow {\hat b} 
\end{equation}
The unitarity of transformation ${ U}$ (equations (\ref{U}) and (\ref{U2})) implies that 
\begin{equation}
\overrightarrow{\hat b^\dagger}^T \cdot \overrightarrow {\hat b} 
= \overrightarrow{\hat a^\dagger}^T {U}^{T *} \cdot  { U}\overrightarrow {\hat a}= \overrightarrow{\hat a^\dagger}^T \cdot  \overrightarrow {\hat a}
\end{equation}
Consequently the total number of photons is an intrinsic operator.

- Let us  define a \textit{"multimode coherent state"} as an eigenstate of all annihilation operators ${\hat b}_m $ in a given mode basis. It is very easy to see that this property is also true in any other mode basis. Therefore this property is intrinsic.

- The mode basis change  is a unitary transformation $U$ that conserves the commutation relations: it is therefore a special case of \textit{symplectic transformation} \cite{Dutta1995}. We will exploit this feature several times in the following. A first consequence is the \textit{invariance of the Wigner function}. More precisely if $W_u(\alpha_1, ..., \alpha_{\ell},...)= W_u(\overrightarrow \alpha)$ is the multimode Wigner function of a given quantum state of light in the phase space of complex coordinates $\alpha_1, ..., \alpha_{\ell},..$ spanning over the $N$ modes $\{{\bm u}_{\ell}\}$ and  $W_f(\overrightarrow \beta )$ is the Wigner function of the same quantum state written now  in the phase space spanning over the modes $\{{\bm f}_{m}\}$, one has \cite{Simon1994}:
\begin{equation}\label{Wevol}
W_f(\overrightarrow \beta) =W_u(\overrightarrow \alpha ) \quad with \quad \overrightarrow \beta = U \overrightarrow \alpha 
\end{equation}
This means that the values of the Wigner function are the same in both bases, but they occur at different values of the coordinates in phase space. In particular \textit{the sign of the Wigner function is intrinsic}. Another intrinsic and additive quantity, related to the volume $N$ of the negative part of the Wigner function is the \textit{Wigner logarithmic negativity}  \cite{Kenfack2004,Albarelli2018}:
\begin{equation}
L_{nw} = \log \left( \int d^N \alpha \, \vert W_u(\overrightarrow \alpha)\vert) \right)
\end{equation}
Therefore the quantum properties related to the fact that the Wigner function of some quantum states have negative parts, are intrinsic. This is the case for the states that yield a "quantum advantage", or are needed for universal quantum computing purposes.
 
A special case of this invariance concerns the value of the Wigner function at the origin, which is related to the parity of the photon number distribution \cite{Royer1977}. One has obviously $W_f(0)=W_u(0)$, which means that \textit{the mean value of the parity operator  is an intrinsic quantity}.

- The same invariance property is also true for the Glauber-Sudarshan $P$ function :
\begin{equation}
P_f(\overrightarrow \beta)=P_u(\overrightarrow \alpha )  
\end{equation}
The sign  of the $P$ function is often related to the non-classicality of the corresponding state \cite{Vogel2000}. Therefore, the \textit{non-classicality of a state is also intrinsic}. Similarly, we have $P_f(0)=P_u(0)$: the probability of being in vacuum state is an intrinsic quantity.

- the purity of a quantum state $P=Tr\rho^2$ can be calculated from its Wigner function
\begin{eqnarray}
P&=&2 \pi \int d^N(\alpha)W_u^2( \overrightarrow \alpha) = 2 \pi \int d^N(\beta)W^2_f(\overrightarrow \beta)
\end{eqnarray}
(as the Jacobian of the coordinate change is 1). As a result, \textit{the purity of a multimode quantum state of light is an intrinsic quantity}.

\section{Search for principal modes}

This section is concerned with the following problem, which is not restricted to quantum physics \cite{Comon1994,Milione2015}: given a quantum multimode state, pure or mixed,  is it possible to find a mode basis which simplifies the expression of the quantum state and reduces it to forms that are more suitable to characterize it physically? The modes of this basis will be called "principal modes". To this purpose, we will use different correlation matrices, in close analogy with the classical case.

More precisely we would like to find a mode basis which allows us to describe a given pure state $|\psi \rangle$ as:
\begin{equation}
|\psi \rangle=|\phi_p\rangle \otimes |0,0...\rangle, 
\end{equation}
or a mixed state $\rho$ as:  
\begin{equation}
\rho=\rho_p \otimes |0,0...\rangle \langle 0,0... |
\end{equation} 
where  $|\phi_p\rangle$ or $\rho_p$ span on a minimal number $p$ of modes ${\bm f}_1, ..., {\bm f}_p$.  $p$ will then be called the \textit{intrinsic number of modes} of the corresponding state.

In this basis, characterized by annihilation operators $ \hat{a}_n$, one has for any $n>p$:
\begin{eqnarray}\label{Nmode}
\hat{a}_n | \psi \rangle=0 \quad &or& \quad  \hat{a}_n  \rho=0  \nonumber\\
 \hat{a}^{\dagger}_m\hat{a}_n | \psi \rangle=0 \quad &or& \quad  \hat{a}^{\dagger}_m\hat{a}_n  \rho=0
\end{eqnarray}

Relations (\ref{Nmode}) imply that the matrix elements $({\bf \Gamma}^{(1)})_{m,n}$ of the coherency matrix such that $m>p$ and $n>p$ are zero. This property is valid for both pure and mixed states. The coherency matrix ${\bf \Gamma}^{(1)}$ consists therefore of a square $ p \times p$ non-zero diagonal sub-matrix surrounded by zeros. Reciprocally, if ${\bf \Gamma}^{(1)}$ has the form we just described, then $\forall n>p \quad \langle \hat{a}^{\dagger}_n \hat{a}_n \rangle=0$: the mean number of photons in $n^{th}$ mode is zero, which implies that the considered state is the vacuum for all modes with $ n>p$. The state is therefore a $p$-mode state, where $p$ is nothing else than the rank of the coherency matrix.

Let us now take a mode basis that we can choose at will, called $\{ {\bf g}_{\ell} \}$, with the associated annihilation operators $\hat{c}_{\ell}$. The corresponding coherency matrix ${\bf \Gamma}^{(1)}=\langle \overrightarrow {{\hat c}^{\dagger}} \overrightarrow{\hat c}^T\rangle$ contains a priori many non-zero matrix elements, so that the quantum state "looks" highly multimode. However, ${\bf \Gamma}^{(1)}$ is Hermitian and therefore diagonalizable. More precisely there is a unitary transformation $V$ that diagonalizes the $N \times N$ matrix, such that:
\begin{equation}\label{cov1}
V {\bf \Gamma}^{(1)}  V^{\dagger} =Diag(n_1, ..., n_p, 0....)
\end{equation}
Let us now introduce the column vector of creation operators in the new basis generated by mode transformation $V$, $\overrightarrow {{\hat d}^{\dagger}}=V \overrightarrow {{\hat c} ^{\dagger}} $. One has:
\begin{equation}\label{cov2}
Diag(n_1, ..., n_p, 0....)= \langle  V \overrightarrow {{\hat c}^{\dagger} }\overrightarrow{\hat c}^T V^{T *} \rangle=\langle   \overrightarrow {{\hat d}^{\dagger} }\overrightarrow{\hat d}^T\rangle
\end{equation}
We have therefore proved that in the mode basis $\{\overrightarrow h \}$ of the modal space defined by  $\overrightarrow h= V \overrightarrow g$, the first order covariance matrix is reduced to the subset of the first $p$ modes, and found the right mode basis and the corresponding minimum mode number.

The diagonalization of the ${\bf \Gamma}^{(1)}$  matrix yields directly the list of the $p$ eigenmodes allowing us to write the quantum state, pure or mixed, in its simplest form. We will see in the following about quantum frequency combs, that the reduction in size of the problem can be drastic, namely from $10^5$ frequency modes to a few principal modes. In particular, it is wise to use the set of principal modes if one wants to make the full tomography of a multimode quantum state, instead of a mode basis with a much larger number of non empty modes. Note that this procedure is valid for Gaussian and non-Gaussian states (see section VII).

The energy content of the principal modes is given by the corresponding eigenvalue, whereas the absence of off-diagonal terms in this basis implies that the principal modes are mutually incoherent: it is not possible to observe interferences on a linear combination of two principal modes. Very often the eigenvalues of the coherency matrix are all non zero, but form a series with decreasing terms. In this case, it is possible to define an effective intrinsic mode number $\bar{p}$, which gives the approximate number of the most excited modes. It can be obtained by the same procedure as for the effective Schmidt number. It is defined by:
\begin{equation}\label{Nbar}
\bar{p}=\frac{\langle \sum_m \hat{N}_m\rangle^2}{\langle \sum_m \hat{N}^2_m \rangle}=\frac{(Tr  {\bf \Gamma}^{(1)})^2}{Tr (({\bf \Gamma}^{(1)})^2)}
\end{equation}
with $\hat{N}_m =\hat {d}^{\dagger}_m \hat{d}_m$.

Note that extracting physically reliable and useful information from the measurement of the noise amplitudes and correlations in a complex physical system by extracting a finite number of "principal modes" from noise matrices is a well-known procedure in other parts of science and technology \cite{Shah2005}. It would be too long to quote all of them. If one restricts oneself to recent developments in optics and electromagnetism \cite{Fan2005}, one can mention the MIMO technique (Multiple In Multiple Out) in telecommunication technologies \cite{Winzer2011} and the control of light propagation in complex media for optical computation purposes \cite{Gigan2017}.

From a given noise matrix experimentally determined by measurements performed on a mode basis appropriate for detection and the computation of principal modes, it is also possible to determine the noise of any physical parameter of the considered system and to get a fruitful physical insight into the underlying physical mechanisms. This method has been in particular recently applied to the analysis of mode-locked lasers \cite{Schmeissner2014}.

\section{Intrinsic single-mode states}

We now focus our attention on the important particular case where the state $\rho$ can be reduced to a single mode state by an appropriate choice of the mode basis, ${\bm f}_m$, with associated annihilation operators ${\hat c}_{m}$. 

Let us first assume that the state we are interested in is the pure state $|\psi \rangle$. Then $|\psi \rangle=|\phi_1\rangle \otimes |0,0...\rangle$ on mode basis ${\bm f}_m$. We know that it has a coherency matrix with a single nonzero eigenvalue, but it can be also characterized by a simpler mathematical property: if one takes a test mode basis ${\bm g}_{\ell}= \sum_p  U^p_{\ell} {\bm f}_p$ with associated annihilation operators ${\hat b}_{\ell}$, one has obviously, for any $\ell$:
\begin{equation}\label{criterion}
 {\hat b}_{\ell}|\psi\rangle= U^1_{\ell} {\hat c}_{1}|\phi_1 \rangle \otimes |0 \rangle
\end{equation}
\textit{In a single mode pure state, all vectors ${\hat b}_{\ell}|\psi\rangle$ in any test mode basis are proportional to each other}. In addition, one can show \cite{Treps2005} that this property is  also a sufficient condition for being single mode.

Let us now assume that the state is in the mixed state $\rho=\rho_1\otimes |0 \rangle \langle 0 |$ on the appropriate mode basis  $\{{\bm f}_m\}$. If one takes a test mode basis $\{{\bm g}_{\ell}\}$, one has, for any $\ell$:
\begin{equation}\label{criterion2}
 {\hat b}_{\ell}\rho=U^1_{\ell} {\hat c}_{1}\rho_1 \otimes |0 \rangle \langle 0 |
\end{equation}
\textit{In a single mode mixed state, all operators ${\hat b}_{\ell}\rho $ are proportional to each other}. Here also this property can be shown to be sufficient for being in a single mode mixed state \cite{Leroyer2007}.

\subsection{Examples}

-a) Let us consider the following state, which is a coherent superposition of single photon states in different modes:
\begin{equation}\label{eq6.5}
|\Psi_1 \rangle = \sum_{m}c_{m}|1: {\bm f}_m \rangle
\end{equation}
with $\sum_m|c_m|^2=1$, where $|1: {\bm f}_m \rangle$ is a short notation for a state with one photon in mode ${ \bm f}_m$ and zero in all the other modes. This state, often called single photon "wavepacket", is an eigenstate of $\hat{N}_{tot}$ with eigenvalue 1. It is actually an intrinsic single mode state as all $\hat{b}_{\ell}|\Psi_1\rangle$ are proportional to the vacuum state: we have therefore showed that \textit{all single photon states are actually single mode states}.

To find the mode in which the single photon can be found, let us introduce the ${\bm v}_1$ mode, given by the same linear combination as (\ref{eq6.5}), but for modes instead of states:
\begin{equation}\label{eq6.6}
{\bm v}_1=\sum_m c_m {\bm f}_m
\end{equation}
It is a first element of a mode basis $\{ {\bm v}_n\}$, that one completes
with orthogonal functions. It is easy to show that in this new
mode basis, the state $|\Psi_1 \rangle$ writes:
\begin{equation}\label{eq6.7}
|\Psi_1 \rangle = |1: {\bm v}_1\rangle \otimes |0,0...\rangle
\end{equation}
${\bm v}_1$ is indeed the mode in which the single photon "lives".

 We have already noticed in a previous section  the equality between the scalar product of single photon states and that of the corresponding modes (relation (\ref{homo})). Such an homomorphism between the mode and state properties for single photons  is perhaps the reason why the detailed consideration of the multimodal structure of quantum states and its physical consequences has not been the object of many studies by physicists interested in single photon states.

\vspace{5mm}
-b) Consider now the \textit{statistical mixture} of single photon states:
\begin{equation}\label{mix1}
\rho_1 = \sum_{m} p_{m}|1: {\bm f}_m \rangle \langle 1: {\bm f}_m |
\end{equation}
with $p_m>0$ and $\sum_m p_m=1$. It is not single mode, because $\hat{b}_{m_1} \rho_1 = p_{m_1} |0\rangle \langle 1:  {\bm f}_m |$ are not proportional to each other. One easily shows that the coherency matrix is diagonal, with eigenvalues which are precisely the statistical coefficients $p_m$.

\vspace{5mm}
-c) Let us now consider the \textit{multimode coherent state}, introduced in section III.E, which is an eigenstate of all annihilation operators ${\hat b}_m $ in a given mode basis. It can therefore be written as a tensor product of coherent states $|\alpha_m:{\bm f}_m \rangle$ of eigenvalue $\alpha_m$ in mode ${\bm f}_m $:
\begin{equation}\label{eq6.8}
|\Psi'_1 \rangle = |\alpha_1: {\bm f}_1\rangle \otimes ..\otimes |\alpha_m:{\bm f}_m \rangle \otimes ..
\end{equation}
This state is single mode as all $\hat{a}_{\ell}|\Psi'_1\rangle$ are proportional to $|\Psi'_1 \rangle$ itself.

To find the mode, let us introduce the ${\bm w}_1$ mode given by:
\begin{equation}\label{eq6.9}
{\bm w}_1=\frac1{\beta}\sum_m \alpha_m {\bm f}_m
\end{equation}
with $|\beta|^2=\sum_m |\alpha_m|^2$. It is the first element
of a new mode basis ${\bm w}_n$, that one completes again. It is here
also easy to show that in this new mode basis, the previous
multimode coherent state is:
\begin{equation}\label{eq6.10}
|\Psi'_1 \rangle = |\beta: {\bm w}_1\rangle \otimes |0,0...\rangle
\end{equation}
Here also, the state $|\Psi'_1 \rangle$, which looks highly multimode in the original basis, is a simple single-mode coherent state in a more adapted basis.

It is easy to show from relation (\ref{comut}) that, for any coherent state, written in the basis where it is single mode in mode ${\bm g}$:
\begin{equation}\label{annihil2}
{\hat b_{{\bm f}}} | \beta: {\bm g} \rangle=  ( \overrightarrow{f}^{T*} \cdot \overrightarrow{g} )\, \beta \, | \beta: {\bm g} \rangle
\end{equation}
where ${\hat b_{{\bm f}}}$ is the annihilation operator in mode $f$. Note that this relation presents close analogies with  the corresponding expression (\ref{annihil}) for single-photon states: the spatio, temporal and polarization properties of coherent states and single photon states are often similar. 

\vspace{5mm}
-d) Consider now the statistical mixture of coherent states:
\begin{equation}\label{mix2}\overline{
}\rho'_1 = \sum_{m} p_{m}|\alpha_m : {\bm f}_m \rangle \langle \alpha_m : {\bm f}_m |
\end{equation}
It is not intrinsically single mode, because the operators $\hat{b}_{m_1}  \rho'_1 $ are not proportional to each other. The coherency matrix is diagonal with eigenvalues $p_m |\alpha_m|^2$.

Examples b) and d) show that mixed states are "more multimode" than the corresponding pure states: the inclusion of classical noise in addition to quantum noise increases the number of principal modes involved in the description of the state. Furthermore, one can describe the statistical properties of such mixed states using the classical coherency matrix, i.e. within the framework of classical coherence theory. While this is not unexpected for coherent states, that are often named "quasi-classical", it is more noteworthy for single photon states.

\vspace{5mm}
-e) Let us consider finally the Hong-Ou-Mandel two-photon pure state $|\Psi" \rangle= |1:  {\bm g_1} \rangle |1:  {\bm g_2} \rangle$. It is a not a single mode state, because $\hat{a}_1|\Psi"\rangle= |0:  {\bm g_1} \rangle |1:  {\bm g_2} \rangle$ and $\hat{a}_2 |\Psi"\rangle= |1:  {\bm g_1} \rangle |0:  {\bm g_2} \rangle$ are orthogonal. This means in particular that there are no mode basis change, or unitary transformation, which will enable us to write it as the single mode state $ |2:  {\bm g_1} \rangle |0:  {\bm g_2} \rangle$: two photons in different modes cannot "merge" into a two-photon state in a single output mode by a unitary transformation.

\subsection{Relation with classical optical coherence}

The notion of optical coherence \cite{Glauber1963,Goodman2015,Mandel1995} is linked to the ability to observe interference fringes with a high visibility. It is characterized by the normalized first order complex correlation function:
\begin{equation}\label{g1}
g^{(1)}({\bf r}, {\bf r}', t, t')=\frac{G^{(1)}({\bf r}, {\bf r}', t, t')}{\sqrt{G^{(1)}({\bf r}, {\bf r}, t, t) G^{(1)}({\bf r}', {\bf r}', t', t')}}
\end{equation}
where $G^{(1)}({\bf r}', {\bf r}', t', t')$ has been defined in (\ref{G1}).  If $|g^{(1)}|=1$ then interferences are of contrast 1, and the field is said to be perfectly coherent.

The quantum counterpart of $G^{(1)}$, where the statistical average represented by the overbar is replaced by a quantum average, is related to the first order coherency matrix that we have introduced in (\ref{compcov}):

For a single mode state in mode ${\bm v}_1$, one has: 
\begin{equation}
\langle \hat{E}_i^{(+)\dagger}({\bf r},t)\hat{E}_j^{(+)}({\bf r}',t')\rangle =
\mathcal{E}_0^2 ({\bf \Gamma}^{(1)})_{1,1}  v^*_{1 i}({\bf r},t)v_{1 j}({\bf r}',t')
\end{equation}
valid for all ${\bf r}, t, {\bf r}', t', i, j$. This implies that $|g^{(1)}|=1$  for all ${\bf r}, t, {\bf r}', t'$. This property does not depend on the quantum state "living" in the mode (except for the energy scaling factor $({\bf \Gamma}^{(1)})_{1,1}=N_{tot}$). A contrario, if there are two points in space time, $({\bf r}_0, t_0)$ and $({\bf r}'_0, t'_0)$ , such that $|g^{(1)}({\bf r}_0, {\bf r}'_0, t_0, t'_0)| < 1$, then one can conclude that the state is not intrinsically single mode.

Consequently, perfect optical coherence is not limited to the use of coherent states: the condition of perfect coherence is related to the modal property of the quantum state, more precisely to the fact that only one electromagnetic mode is not in the vacuum state. It does not depend on the properties of the quantum state defined in this mode. For example, one will be able to see perfect interference fringes with coherent states, which is not unexpected, but also with strongly "incoherent-looking" states such as single photon states, as noticed by Glauber \cite{Glauber1963}, or with strongly "quantum-looking" states such as Schrodinger cats. In contrast, this will not be possible with the Hong-Ou-Mandel two-photon state $|1,1\rangle$ which is not single mode.

\subsection{Simple sufficient conditions for an intrinsic single mode state}

The single-mode criteria that we have just exposed are not easy to implement experimentally. One possibility is of course to experimentally measure all the matrix elements of the quadrature covariance matrix, which will be explained in section X B2. As this is not a simple task, it would be obviously interesting to have at one's disposal a sufficient physical criterion that can be experimentally checked in a simpler way. 

The physical meaning of an intrinsic single mode state is simple: in such a state, all the physical properties have a spatio-temporal dependence given by ${\bm v}_1$. As a result the mean value $<\hat{E}^{(+)}(x,y,z,t)>$ and the standard deviation $\Delta{E}^{(+)}(x,y,z,t)$ have the same spatial dependence. In contrast, in non-intrinsic single-mode states, these different quantities may have different spatial variations. So a rather easy experimental check of whether a given quantum state is intrinsically single mode or not, consists of measuring the mean value and the quadrature noise in different areas, in different time windows or in different spectral regions. If the ratio of these two quantities is not constant when one varies the detection area, the time bin or the frequency band, then one is sure that the state is not intrinsically single mode. Note that this test gives only a  sufficient condition: a constant value of the ratio does not imply that the state is single mode. This criterion has been successfully used by different groups in experiments involving spectral  \cite{Spalter1998,Marin1995} and spatial modes \cite{Hermier1999,Martinelli2003,Corzo2011}. For example in  \cite{Marin1995}, the intensity noise of a sub-Poissonian laser diode is spectrally resolved using a spectrometer of variable width. The nonlinear variation of the intensity noise as a function of the number of spectral modes of the diode laser is an indication of the existence of strong anti-correlations between the main longitudinal mode and the weak side modes.

\section{Counting and determining the principal modes}

Let us now consider more complex quantum states than the intrinsic single mode ones. The minimum number $p$ of modes that are needed to completely describe them, and the shape of the corresponding principal modes are important parameters that characterize  the system under study. From a mathematical point of view,  it can be shown in the case of a pure state that $p$ is also the dimension of the vector space generated by all the vectors $\hat{b}_{m}|\Psi\rangle$, $\{ \hat{b}_{m} \}$ being the set of annihilation operators associated with a given test mode basis. From a more practical point of view, the best way is to experimentally determine all the the matrix elements of the coherency matrix, to diagonalize it and to count the number of non-zero eigenvalues \cite{Morin2013}. This method works for pure and mixed states, and for Gaussian and non-Gaussian states as well.

Counting modes is of course not restricted to quantum optics, and researchers have developed several criteria that allow them to find the number of such modes in various situations. It is therefore interesting to compare the different ways of determining the mode number in a  multimode system.

\subsection{Case of spatial modes}

We will first treat in this section the case of $1D$ spatial modes. They can be  readily extended to $2D$ spatial modes. In classical optics, the number $p$ of spatial modes in a light beam of transverse size $D$ is usually taken as equal to  $D/d_c$, where $d_c$ is the coherence length \cite{Karny1983}. To compare this value with the one determined in section IV we need to find a quantum state $\rho$ describing the same physical situation. A possible state is the following statistical mixture:
\begin{equation}
\rho= \sum_{n=1}^p p_n |\alpha_n :  \bm{w}_n \rangle \langle \alpha_n : \bm{w}_n|
\end{equation}
with $\sum p_n  =1$. The modes $\bm{w}_n(x)$ are pixel modes for which $w_n(x)$ is equal to $1/\sqrt{d_c}$ for $(n-1) d_c <x< n d_c$ and zero outside. Their number is precisely $p$.  $ |\alpha_n  :  \bm{w}_n  \rangle$ are coherent states defined in each coherence area. Let us now apply our quantum counting technique to this state: in such a quantum state the first order coherency matrix is diagonal in the pixel mode basis because the fluctuations in the different pixel modes are uncorrelated. Its diagonal elements are $p_n |\langle \alpha_n |\hat {a}^{\dagger}_n \hat {a}_n |\alpha_n\rangle |^2$: the number of non zero diagonal elements is indeed $D/d_c$, so that the  classical and quantum approaches lead to the same value of the mode number.

We will consider in Appendix A other multimode light sources, such as lasers and parametric down conversion devices. We will show on a simple example than the number $p$ of transverse modes in a laser is equal to the $M$ factor introduced by Siegman \cite{Siegman1998}, and that it is equal to twice the Schmidt number in the case of parametric down conversion.

On the experimental side, the spatial structure of the mode of an heralded single photon has been recently experimentally determined by a holographic technique \cite{Chrapkiewicz2016}.

\subsection{Case of temporal modes}

The generation of pure single photons is the object of an intense experimental research \cite{Lvovsky2001,Eisaman2011}. We have seen that, theoretically speaking, pure single photons are always single mode. Experimentally speaking, more than one mode can be populated because of experimental imperfections. It is therefore very important to evaluate the number and the exact shape of the temporal modes that are excited.
This problem has been for example theoretically considered in \cite{Aichele2002}. A convenient way to do it is to derive it from the information contained in the temporal coherency matrix, also named in this context the temporal autocorrelation function \cite{Nielsen2007}, the matrix elements of which are:
\begin{equation}\label{compcov2}
({\bf \Gamma}^{(1)} )_{n,n'}= \langle {\hat a}^{\dagger}(t_n) {\hat a}(t_{n'} )\rangle
\end{equation}
where $t_n$ and $t_{n'}$ are sampling times of the homodyne signal, taken in large enough quantities to reach a good statistical estimation. The diagonalization of this matrix allows us to know the temporal shapes, and the degree of excitation, of the  different principal modes.

Let us present briefly the method used in \cite{Morin2013}, one of the first fully experimental determinations of the temporal modes of several heralded states, namely single-photon, two-photon, and coherent state superpositions: one produces by type II parametric down conversion weak signal and idler beams.  The temporal fluctuations of the signal beam quadrature component are continuously recorded by homodyne detection. When a photon is detected at time $t_c$ on the idler beam, one post-selects  the homodyne signal values in $N/2$ time bins before $t_c$ and $N/2$ time bins after $t_c$. These data are then averaged over many detected idler photons and used to determine the $N \times N$ temporal auto-correlation function. If the phase of the Local Oscillator is random during the accumulation time, it is easy to show that the experimental autocorrelation matrix is directly related to the coherency matrix in the time bin temporal mode basis. Its diagonalization gives the number and the temporal shape of the modes which are not in the vacuum state. If this state is indeed single photon, the matrix has a single eigenstate with a non-zero eigenvalue. Its temporal shape on the $N$ time bins gives the shape of the mode in which "lives" the conditionally generated single photon. A similar approach has been used in \cite{Qin2015}.

Other recently developed technique use spectral shearing interferometry \cite{Davis2018} or dual homodyne measurement \cite{Takase2019}.  Sum frequency generation can also be used as a kind of fast correlator which allows also to determine the number of modes in parametric down conversion  \cite{Kopylov2019}.

\section{Multimode Gaussian states}
 
Gaussian states play a very important role in Continuous Variable quantum optics, as they are non-classical states that are deterministically generated by nonlinear optical processes in which the quantum fluctuations are small compared to the mean values. They have also the practical interest of being completely characterized by the first and second moments of the quadratures (but also of the photon numbers \cite{Vallone2019}). In addition, they are promising candidates as building bricks for Quantum Information and Quantum Metrology purposes \cite{Adesso2014,Weedbrook2012} .

\subsection{Symplectic transformations}

Let us use again the compact vectorial notation $\hat {\overrightarrow Q}$ defined in (\ref{Q}) containing the quadrature operators of all  $N$ modes. The usual commutation relations can then be written as
\begin{equation} 
 [\hat{Q}_\mu, \hat{Q}_\nu] = 2i\beta_{\mu,\nu} \quad \textrm{where} \quad
\beta = \left(
\begin{array}{cc}
\mathbb{0} & \mathbb{1} \\
\mathbb{-1} & \mathbb{0}
\end{array}\right).
\end{equation}

A symplectic transformation is defined by a $2N \times 2N$ real matrix  $S$ acting on the quadrature operator column vector  $\hat{\overrightarrow Q}$ that preserves the commutation relations \cite{Simon1988}. The set of such transformations is named the Symplectic Group (\cite{Dutta1995}). The condition that commutators are preserved leads to the following condition for a  symplectic transformation $S$: 
\begin{equation}
S\beta S = \beta
\end{equation}

This group includes obviously all the mode basis change transformations that we have already considered, but it is not limited to these transformations.

\subsection{Wigner function of a Gaussian state}

By definition, a Gaussian state is a state, pure or mixed, which is described by a Gaussian Wigner function. It is, modulo a classical displacement, completely described by its covariance matrix  ${\bf \Gamma}_Q$. Note that all positive matrices cannot be used as the covariance matrix of a Gaussian state, as they need to satisfy the generalized Heisenberg inequality (\ref{ineq}). In addition  the covariance matrix ${\bf \Gamma}_Q$ associated with a non-classical (squeezed and/or entangled) quantum state has at least its smallest eigenvalue which is smaller than 1.

The Wigner function is a real quasi-probability distribution over multimode phase space depending on the $2N$ real quadrature  coordinates $\overrightarrow q $. In the case of a Gaussian state of covariance matrix  ${\bf \Gamma}_Q$ its expression is
\begin{equation}\label{WignerGaussianMultimode}
W (\overrightarrow q) = \frac{1}{(2\pi)^N\sqrt{\det{\bf \Gamma}_Q}}e^{-(\overrightarrow q-\langle \overrightarrow Q \rangle)^T{\bf \Gamma}_Q^{-1}(\overrightarrow q-\langle \overrightarrow Q \rangle)}.
\end{equation} 

The purity $P$ of a Gaussian state is simply related to the covariance matrix by the relation:
\begin{equation}\label{purity}
P=\frac1{\det{\bf \Gamma}_Q}
\end{equation} 

The transformation of the Wigner function under a symplectic transformation $S$ can be directly calculated using the same approach as in (\ref{Wevol}):
\begin{equation}
W'(\overrightarrow q') = W(\overrightarrow q) \quad \textrm{with} \quad \overrightarrow q' = S\overrightarrow q
\end{equation}
Hence a Gaussian state remains Gaussian under symplectic transformation. Moreover the values of the Wigner function are conserved in the transformation (and in particular the negative ones) but found at different points of the phase space. The covariance matrix change under the effect of a symplectic transformation $S$ is expressed as:
\begin{equation}\label{CovaS}
{\bf \Gamma}_Q' = S {\bf \Gamma}_Q S^T
\end{equation}

\subsection{Gaussian state decomposition}

\subsubsection{Bloch Messiah reduction}\label{sec:BM}
Many decompositions of symplectic transformations on subgroups of the symplectic group do exist, and we refer the reader to \cite{Dutta1995} on this subject. Within the context of quantum optics, the most relevant one, introduced by \cite{Braunstein2005} is the so-called Bloch-Messiah reduction \cite{Bloch1962}. It is a direct consequence of the singular value decomposition. 

Any symplectic $S$ matrix can be decomposed into three matrices such as:
\begin{equation}\label{BM}
S = O_1 K O_2
\end{equation}
where $O_1$ and $O_2$ correspond to mode basis changes. In equation (\ref{U}), a mode basis change was characterized by a unitary modal transformation $U$ acting on the complex creation operators. The same mode basis change is also characterized by the following real orthogonal matrix $O$ acting in the real space of quadrature operators:
\begin{equation} \label{OrthogonalO}
O = \left(
\begin{array}{cc}
\textrm{Re}(U) & \textrm{Im}(U) \\
-\textrm{Im}(U) & \textrm{Re}(U)
\end{array}\right).
\end{equation}
where  $U = \textrm{Re}(U) + i\textrm{Im}(U)$. $K$ is a diagonal matrix of the form $K = diag(\sigma_1, \sigma_2, \ldots, \sigma_N, \sigma_1^{-1}, \sigma_2^{-1}, \ldots, \sigma_N^{-1})$, where $\sigma_i$ are real and positive numbers. It corresponds to a multimode squeezing operation in a well-defined mode basis.

\subsubsection{Williamson reduction}

The evolution of the covariance matrix under symplectic transformations as expressed in (\ref{CovaS}) is not a usual diagonalisation procedure, and thus the standard spectral theorem does not apply. One can show \cite{Dutta1995} that a diagonal form of the covariance matrix can always be found under these transformations, process known as the Williamson reduction, that we expose here. For any covariance matrix ${\bf \Gamma_Q}$ associated with a physical quantum state it exists a symplectic transformation $S'$ such that:
\begin{equation} \label{Williamson}
S'{\bf \Gamma}_Q S'^T = {\bf \Gamma}_W = Diag(\kappa_1, \ldots, \kappa_n,\kappa_1, \ldots, \kappa_n) 
\end{equation}
where $ 1\leq \kappa_1 \ldots \leq \kappa_n $ are named the Williamson eigenvalues. This means that under symplectic transformation, any Gaussian state can be reduced into a collection of independent symmetric thermal states with $\langle {\hat X}_i^2\rangle= \langle {\hat P}_i^2\rangle=\kappa_i$. Furthermore, because state purity is invariant under symplectic transformation one has that $P = 1/\Pi_i \kappa_i$.

Combining Williamson and Bloch Messiah reductions (\ref{BM}) and (\ref{Williamson}), one finds that for any Gaussian state
\begin{equation}\label{MixedBlochMessiah}
{\bf \Gamma}_Q = O_1KO_2 {\bf \Gamma}_W O_2^T K O_1^T
\end{equation}
where ${\bf \Gamma}_W$ is of the form introduced in (\ref{Williamson}).

\subsubsection{Pure Gaussian state}
For a pure Gaussian state, ${\bf \Gamma}_W$ is the identity matrix. Hence one can write:
\begin{equation} \label{PureBlochMessiah}
{\bf \Gamma}_Q^{pure} = O_1K^2O_1^T.
\end{equation}
Any pure Gaussian state can then be seen as the result of a mode basis change on a collection of independent pure squeezed states, thus inducing quantum entanglement between the modes \cite{Braunstein2005}. This remarkable property is used for instance to construct complex multimode entangled Gaussian states from a set of independent squeezers and a generalized interferometer (see section XII A)

\subsubsection{Mixed Gaussian state}
For general Gaussian mixed states (\ref{MixedBlochMessiah}) cannot be simplified, hence the second basis change $O_2$ is now physically relevant.  Relation (\ref{MixedBlochMessiah}) has a clear physical meaning: any Gaussian state can be constructed from a generalised factorised thermal state ${\bf \Gamma}_W$ with a first basis change $O_2$ which induces classical correlations between the input modes,  which are then fed into a multimode squeezing operation $K$ and finally mixed again by another basis change $O_1$. 

The basis changes and the multimode squeezing characteristics can be straightforwardly deduced from the measured covariance matrix. Such a procedure enables us to uncover the modes carrying the quantum properties and those carrying the classical noise. However, one should be careful in associating too much "quantumness" to the $K$ matrix when it acts on a thermal state.  In this case, the multimode squeezing matrix modifies in a phase dependent way the input classical noise, and the resulting output multimode mixed state can still be classical.

\subsubsection{Intrinsic separability}\label{intrinsic-separability}
The notion of separability, which corresponds to the ability to write a quantum state as a statistical superposition of factorised states \cite{Peres1996,Li2008}, is inherently mode basis dependent and we will review its counterpart, entanglement, in the next section. Separability conditions can be derived from the covariance matrix \cite{Gittsovich2008}. The underlying symplectic geometry of Gaussian states renders possible the study of {\it intrinsic separability}, i.e. the question of the possible existence of a mode basis in which a given multimode state is separable \cite{Mancini2006}. For pure Gaussian states, this is a direct consequence of Bloch Messiah decomposition, and the basis in which a multimode gaussian states is separable is given by eq.(\ref{PureBlochMessiah}). For mixed Gaussian states eq.(\ref{MixedBlochMessiah}) does not provides directly the solution, as the symplectic transformation involved is not a basis change. We thus follow here the approach of \cite{Walschaers2017} to demonstrate the intrinsic separability of any Gaussian state.

One can simply rewrite eq.(\ref{MixedBlochMessiah}) introducing ${\bf \Gamma}_{th} = O_2 {\bf \Gamma}_W O_2^T$, which is the covariance matrix of a thermal state. A Gaussian state can thus be decomposed as ${\bf \Gamma}_G = {\bf \Gamma}_s + {\bf \Gamma}_c$, where ${\bf \Gamma}_s= O_1K^2O_1^T$ is a pure multimode squeezed vacuum and ${\bf \Gamma}_c = O_1({\bf \Gamma}_{th}-1)O_1^T$  corresponds to the added noise. Note that ${\bf \Gamma}_c$ is a positive symmetric matrix  that does not in general satisfy Heisenberg inequality (\ref{ineq}) and therefore cannot be associated with a quantum state: it can be seen as a noisy channel randomly displacing the pure state. One can show that the Wigner function of the Gaussian mixed state can be written as the convolution product \cite{Walschaers2017}:
\begin{equation}\label{eq:GaussianMix}
W_G(\overrightarrow q) = \int {\rm d}^{2N}\xi \, W_s(\overrightarrow q-\overrightarrow\xi) p_c(\overrightarrow \xi),
\end{equation}
where 
\begin{equation}
p_c(\overrightarrow \xi) = \frac{e^{-\overrightarrow \xi^T {\bf \Gamma}_c^{-1 }\overrightarrow \xi}}{(2\pi)^N \sqrt{\det {\bf \Gamma}_c}},
\end{equation}
is a Gaussian positive and normalized probability distribution and 
\begin{equation}
W_s(\overrightarrow q) = \frac{e^{-\overrightarrow q^T {\bf \Gamma}^{-1}_s \overrightarrow q}}{(2\pi)^N\sqrt{\det {\bf \Gamma}_s}}. 
\end{equation}
is the Wigner function of a factorized multimode squeezed state. Expression (\ref{eq:GaussianMix}) is nothing but the statistical mixture of displaced pure squeezed vacuum states, all separable in the same basis defined by transformation $O_1$. Hence, for any Gaussian state (the demonstration here can be easily extended to displaced states) one can find a mode basis, given by Bloch-Messiah-Williamson reduction, in which it is separable.  

Let us stress here this important property of multimode Gaussian states: they are all {\it intrinsically separable}, meaning that they can always be "disentangled" in some appropriate mode basis. Note that this basis is not unique.

\section{Multipartite entanglement}\label{sec:multipartiteentanglement}

Entanglement \cite{Schrodinger1935} and non-separability \cite{Werner1989} are basic quantum resources. Their characterization is subtle and still the object of numerous investigations. These have mostly concerned bipartite systems and entanglement criteria have been introduced \cite{Horodecki2009,Guhne2009} in terms of Schmidt number, Partial Transpose, variances of combinations of quadratures, quadrature matrix eigenvalues ... The complexity of the separability problem increases substantially when one studies multipartite systems. In these situations, one has a rapidly increasing number of
choices in the bunching of parties on which one searches for a possible factorization \cite{VanLoock2000}. Without going into much detail, we present now a brief overview of the domain, restricting ourselves mostly to results which are  scalable to an arbitrary number of partitions. 

In addition of being more complex than bipartite,  multipartite entanglement has some specific features:
\begin{itemize}
\item
Whereas the Schmidt decomposition is a very useful tool for pure bipartite states, there is no such simple decomposition in the case of more than two parties in the general case \cite{Pati2000,Acin2000}.

\item The $N$ multimode system can be divided in many different bi-partitions, but there are also numerous possibilities of multipartitions in $K=3, 4, ...,N$ parties. For example for $N=10$, the total number of multipartitions, whatever $K$, is 115974!  A given state can be entangled for some partitions, not for others, which gives rise to a complex topology of quantum correlations \cite{Menicucci2006}. An important notion is that of "genuine entanglement":  one defines  a \textit{genuinely K-entangled state} as a state that is not a statistical mixture of K-partite factorized density matrices. Genuine entanglement implies multipartite entanglement for every other partition of the modes. However, if a state does not exhibit this specific kind of entanglement (i.e., is two-separable), no conclusions on other forms of multipartite quantum correlations can be drawn.

\item There are in some cases relations between the entanglement measures of the different partitions, which are termed under the name of \textit{monogamy}: in the case of three parties A, B and C, for example, it has been shown,  in the particular cases of qubits \cite{Coffman2000} and symmetric gaussian states \cite{Adesso2006} that the A-BC entanglement between A and the two other modes is stronger than the sum of the "partial" entanglements A-B and A-C. This property is specific to quantum entanglement, in opposition to the classical correlations, which are not constrained and can be freely shared. This property can be generalized to an arbitrary number of bipartitions.

\end{itemize}

Multimode entanglement is also present in the case of a bipartition into two parties which are themselves multimode. This is the case for example of parametric down-conversion (see section IX A). If the two-party quantum state is pure its Schmidt decomposition gives the principal Schmidt modes of the system, and the number of terms (Schmidt number) gives the dimensionality of the system \cite{Law2004,Gatti2012}. In the general case, as the Schmidt decomposition cannot be extended to more than 2 parties, a "vector" of Schmidt numbers for all bipartitions is useful to characterize multipartite pure states \cite{Huber2013}. One can also use in this case a necessary and sufficient entanglement criterion \cite{Gessner2016}.

The case of a mixed state $\rho$ is more difficult to deal with. Theoreticians have introduced various entanglement witnesses for multipartite entanglement, i.e. operators $\hat W$ such that $Tr(\rho \hat W)<0$ implies non separability. There are simple ones, easy to calculate \cite{Hillery2010}, but "missing" many entangled states, and optimized ones \cite{Hyllus2006}. The method of separability eigenvalues \cite{Sperling2013} has been used to characterize all the partitions of a multipartite-entangled quantum frequency comb \cite{Gerke2015}. In addition to witnesses, one can use in the multipartite case the Partial Transpose method \cite{Simon2000}, and define a measure of entanglement in the Continuous Variable regime in terms of logarithmic negativity \cite{Adesso2004}. One can also derive simple criteria to detect genuine entanglement \cite{Toscano2015}. Some states appear to be not genuinely entangled, and therefore two-separable, and yet exhibit a rich multipartite entanglement structure for multipartitions in more than two-parties \cite{Gerke2016}.

The particular case of Gaussian multimode states has been thoroughly investigated with the help of symplectic group methods \cite{Adesso2007,Adesso2012,Franke2013} and criteria based on symplectic invariants \cite{Serafini2006}. Whereas for bipartite Gaussian states, the Partial Transpose  criterion is a necessary and sufficient entanglement identifier \cite{Peres1996}  this is no longer the case for more than two parties: there are multipartite Gaussian states  whose entanglement cannot be uncovered by the partial transposition \cite{Werner2001, Diguglielmo2011}. The logarithmic negativity, a measure of entanglement, can be directly calculated from the covariance matrix, as well as the appropriate measure $E$ that allows us to test the monogamy inequality for Continuous variables \cite{Coffman2000}. In addition the difference $E(A-BC)-E( A-B) - E(A-C)$ gives some information about higher order entanglement \cite{Adesso2007a}. Multipartite steering can be also defined, calculated and used in a monogamy inequality \cite{Xiang2017}. The best way to study the detailed structure of entanglement in a given multimode state is to use criteria which are an extension of  the well-known  Duan criterion \cite{Duan2000}, detailed in \cite{Teh2014,VanLoock2000,VanLoock2003,Toscano2015} in terms of combinations of different quadrature operators, for example ${\hat X}_1 - \sum_i g_i {\hat P}_i  $ and  ${\hat P}_1 + \sum_i g_i {\hat X}_i$ . Reconfigurable Gaussian entangled states have been experimentally produced, characterized and used for quantum information purposes \cite{Yokoyama2013,VanLoock2003,Cai2017,Menicucci2008,Chen2014,Titchener2016}.

Multipartite entanglement of non Gaussian states has been also studied, for example in terms of Mandel matrices that involve normally ordered 4th order correlations \cite{Ivan2011}. Sufficient conditions for genuine multipartite Gaussian and non Gaussian states  have been derived \cite{Shchukin2015}. For photon-added or subtracted Gaussian states, which are the most studied CV non Gaussian states, a hierarchy of inseparability criteria can also been used to characterize precisely the entangled state \cite{Valido2014,Levi2013}, and the relations between the negativity of the Wigner function, purity and entanglement \cite{Walschaers2017}, involving also higher statistical moments of the quadrature operators. The effect of mode-selective photon addition and subtraction on the propagation of entanglement over the quantum network has been theoretically studied in \cite{Walschaers2018} and experimentally investigated  \cite{Ra2019}. It turns out that the non-Gaussian character induced by photon subtraction spreads not further than to the next-to-nearest neighbor node in the cluster graph.

\section{Sources of multimode nonclassical states}

In this section, in order to avoid cumbersome presentation, we will essentially restrict ourselves to the continuous variable aspects of multimode nonclassical state generation.

\subsection{Mixture of single mode non-classical states}

In order to generate any N-mode Gaussian state, a possibility is to start from a bunch of independent single mode squeezed states followed by a linear N-port interferometer comprising beamsplitters and phase shifters \cite{Braunstein2005}. This is true in particular for generating cluster states \cite{Zhang2006}.  It has been shown \cite{Reck1994} that actually any modal unitary operator can be constructed using a sequence of beam splitters transformations.

Multipartite entanglement and quantum networks can be created by this technique using a single squeezed state \cite{VanLoock2000} or several \cite{Armstrong2012,VanLoock2007,Su2014}. Integrated optics and multimode fibers have been also used \cite{Mohanty2017}, as well as spatially multiplexed detection of up to 8 different spatial modes \cite{Armstrong2012,Su2012}. In the pulsed regime time delays can also be used to mix different pulses in the same beam \cite{Yokoyama2013}.  Because single mode squeezers are now capable of reaching impressive amounts of squeezing, up to 15dB \cite{Vahlbruch2016}, this technique is attractive from an experimental point of view. Experimental implementations become more and more complex as $N$ increases and lack flexibility because they are not simply reconfigurable.

\subsection{Parametric down-conversion}

\subsubsection{Interaction hamiltonian}

The most widely employed technique to directly generate, with a single device,  multimode non-classical states is to use twin photon generation by parametric down conversion in a  nonlinear $\chi^{(2)}$ crystal \cite{Grynberg2010}-chapter 7. Using a discrete mode basis and assuming undepleted pump, it is described by the following interaction Hamiltonian:
\begin{equation}\label{param}
\Hat H = \sum_{\ell, \ell'} ( G_{\ell, \ell'} \hat a_{\ell}^{\dagger}  \hat a_{\ell'}^{\dagger} \, + G_{\ell, \ell'}^* \hat a_{\ell}  \hat a_{\ell'})
\end{equation}
where $\hat a_{\ell}^{\dagger}$ and $ \hat a_{\ell'}^{\dagger} $ are creation operators of photons in  modes ${\bm f}_{\ell}$ and ${\bm f}_{\ell'}$ of respective frequencies $\omega_{\ell}$ and $\omega_{ \ell'}$. $G_{\ell, \ell'}=G_{\ell', \ell}=a_{\ell, \ell'} \alpha_p(\omega_{\ell}+\omega_{ \ell'})$, where $ \alpha_p(\omega)$  is the pump field amplitude at frequency $\omega$, and $a_{\ell, \ell'}$ is a coefficient which depends on the nonlinear medium and the geometry of the interaction.

The Hamiltonian is a double sum of EPR entangling $\hat a_{\ell}^{\dagger}  \hat a_{\ell'}^{\dagger}$ terms, so one expects entanglement to be generated between all pairs of twin modes for which the joint two-photon matrix $G$ has significant matrix elements $G_{\ell, \ell'}$. Indeed in the weak pump approximation ($G_{\ell, \ell'} \ll 1$) and vacuum state input, the state at the output of the crystal will be the entangled highly multimode twin-photon state:
\begin{equation}\label{twin}
|\Psi_{out}\rangle= |0\rangle -i\frac{L}{\hbar c} \sum_{\ell, \ell'} G_{\ell, \ell'} |1: {\bm f}_{\ell}\rangle \otimes |1:{\bm f}_{\ell'}\rangle
\end{equation}
where $L$ is the crystal length.

$\quad$ a) If the symmetrical matrix $G$ is real, one can diagonalize it by an orthogonal transformation $O$, which is a special case of mode basis change:
\begin{equation}\label{real}
O G O^{-1} = \Lambda
\end{equation}
where $\Lambda$ is a real diagonal matrix of eigenvalues $\lambda_i$. So in the eigenmode basis with  annihilation operators ${\overrightarrow {\hat b}}=O^{T} {\overrightarrow {\hat a}}$,  the hamiltonian writes:
\begin{equation}\label{supermode}
\Hat H = \sum_{i} ( \lambda_{i} \hat b_{i}^2 \, + \, H. \, C. )
\end{equation}
where the eigenvalues $ \lambda_{i}$ are real.  The eigenmodes are often called "supermodes".

$\quad$ b) If $G$ is complex, using the Autonne-Takagi factorization method \cite{Cariolaro2016}, one can find a unitary matrix $U$ (hence a mode basis transformation) such that 
\begin{equation}\label{takagi}
U G U^T = \Lambda
\end{equation}
 $\Lambda$ being again a diagonal real non-negative matrix. Using the mode transformation ${\overrightarrow {\hat b}}=U^{\dagger} {\overrightarrow {\hat a}}$, the Hamiltonian can be written in the new basis as in (\ref{supermode}). 
 
 In both cases $\Hat H$  is in the new basis a sum of squeezing hamiltonians, which means that the propagation of an initial vacuum state in the nonlinear crystal will lead to a final quantum state which is a tensor product of vacuum squeezed states, the variance in dB of the squeezed quadrature $\hat {X}_i$ in mode $i$ being proportional to the eigenvalue 
$\lambda_{i} $ \cite{Lvovsky2005,Wasilewski2006,Valcarcel2012}. The number of non-zero eigenvalues (i.e. the rank of the $G$ matrix) gives the intrinsic number of non-vacuum modes. It turns out that this number roughly equal to to the aspect ratio of the   $G_{\ell, \ell'}$ matrix (ratio of its widths in the $\omega_{\ell}+\omega_{\ell'}$ and $\omega_{\ell}- \omega_{\ell'}$ directions).

 $\quad$ c) In some instances, like in type II phase matching for example, the $2N \times 2N$ joint two-photon matrix $G$ can be written in terms of $N \times N$ block matrices of the form:
 \begin{equation}\label{Gsignalidler}
 G=\left( \begin{array}{cc}
\mathbb{0} &  G_{si} \\
  G^*_{si} & \mathbb{0}
\end{array}\right)
\end{equation}
The interaction Hamiltonian is a sum of terms creating one photon in one of the  signal modes ${\bm f}^s_i$ and one photon in one of the idler modes ${\bm f}^i_j$. The singular value decomposition of matrix $G_{si}$ writes
\begin{equation}
U_s G_{si} U^{T}_i = \Lambda_{si}
\end{equation}
where $U_s$ and $U_i$  are unitary mode basis changes in the signal and idler parts of the modal space and  $\Lambda_{si}$ the real diagonal matrix $diag(\lambda_1, ... , \lambda_N)$. The singular value decomposition thus generates two sets of  $N$ eigenmodes,  $\{{\bm f}^s_k\}$ and $\{ {\bm f}^i_k\}$, which are EPR entangled with each other. They are the Schmidt modes \cite{Sharapova2018} already introduced in section (VI D). They form a mode basis respectively in the signal and idler parts. One can show \cite{Horoshko2019} that the eigenvalues of the joint two-photon matrix $G$ are \textit{doubly degenerate}, with eigenvectors  $({\bm f}^s_j \pm {\bm f}^i_j )/\sqrt2$, which are therefore both equally squeezed and mutually uncorrelated, with a squeezing factor in dB proportional to $\lambda_j$ . The number of non-zero terms in the Schmidt decomposition gives the intrinsic dimension of the generated multimode state (see appendix B).

\subsubsection{Different possible modes for entangled states}

The entanglement generated by parametric down-conversion concerns different kinds of modes: 
\begin{itemize}

\item polarization modes, that we do not consider in this review;

\item spatial modes \cite{Law2004,Jedrkiewicz2004,Walborn2010}, which can be the transverse modes of cavities \cite{Kolobov2006} or of multimode optical fibers \cite{Jachura2014}.  This leads to non-classical correlations between different parts of an optical image, or between two images;

\item time/frequency modes \cite{Shifeng2012}, which leads to correlations between different spectral components of the light source;

\item polarization, spatial and/or temporal degrees of freedom at the same time, for example polarization/transverse modes \cite{Gabriel2011,Jedrkiewicz2012}, or spectral/temporal/transverse modes \cite{Perina2016}.

\end{itemize}

Nonlinear effects in nonlinear crystals are usually weak and generate  twin-photon states of the form (\ref{twin}). If one uses a strong pulsed pump \cite{Sharapova2018} multimode pulsed twin beams containing more than $6.10^5$ photons have been generated and used to conditionally generate sub-Poissonian light \cite{Iskhakov2016}. In order to enhance the effect, it is possible to use resonant cavities. We will detail this configuration in the following section.

\subsubsection{Use of resonant optical cavities}

Another possibility is to insert the crystal in a resonant cavity. The device is then an Optical Parametric Oscillator (OPO), which produces above some pump threshold bright output beams in the twin modes $\ell$ and $\ell'$ that are resonant with the cavity. As cavities are actually mode filters, there is often only one couple of twin modes that has such a resonant property, and the OPO generates below threshold a two-mode EPR state, or a squeezed vacuum single mode state if the twin photons are generated in the same mode  $\ell$. The squeezing,  or the entanglement, becomes in principle perfect when one approaches the oscillation threshold from below. 

To produce multimode non-classical states of larger dimensionality \cite{Lugiato1993}, one can use a cavity which simultaneously resonates on several couples of parametrically generated modes. Below threshold, the generated quantum state is a tensor product of squeezed vacuum states in each "supermode" $i$, with a squeezed quadrature noise equal to \cite{Patera2009}
\begin{equation}
\Delta X_i^2=\left( \frac{\lambda_1-r |\lambda_i |}{\lambda_1 +r |\lambda_i|}\right)^2
\end{equation} 
where $r$ is the pump field amplitude normalized to the pump amplitude at threshold. Hence the squeezing in the first supermode becomes perfect when one approaches the threshold from below. All the other modes of smaller but nonzero eigenvalue $\lambda_i$ are also squeezed, but by smaller amounts and never get perfectly squeezed. 

If one uses cavities with spherical mirrors, the spatial eigenmodes of which are the Hermite-Gauss modes $TEM_{pq}$, cylindrical symmetry leads to $p+q$ mode degeneracy, that has been for example exploited in \cite{Marte1998,Lugiato2005}. These modes are entangled by the parametric interaction \cite{Schwob1998,Chalopin2011,Lassen2007}. Confocal cavities have a stronger degeneracy and resonate for any spatial mode of symmetrical shape \cite{Martinelli2003}. Self-imaging cavities are resonant for any transverse shape of the electric field \cite{Lopez2009}. 

Cavities are also filters in the spectral domain: they have equally spaced resonant frequencies and enhance not only resonant single frequency modes, but also frequency combs: this leaves room for a possible great number (more than $10^5$ in realistic experimental conditions) of entangled frequency modes. 

\subsubsection{Use of pump modes of different shapes}

In most experiments, the parametric medium is pumped by a monochromatic field, so that the entangled signal and idler modes have to fulfil the relation $\omega_{\ell}+ \omega_{\ell'}=\omega_{pump}$: one gets a set of independently entangled couples of signal and idler modes, but without any multipartite entanglement. The use of a bi-chromatic pump \cite{Chen2014} allows physicists to greatly enhance, and in a well controlled way, the number (up to 60) and the topology of entangled couples of modes: a whole zoology of cluster states \cite{Pfister2011,Pysher2011} can thus be generated, with applications to measurement based quantum computation \cite{Menicucci2008}. 

One may also use as a pump field an optical frequency comb with the same frequency spacing as the OPO cavity (synchronously pumped OPO or SPOPO). If for example the pump spectrum has a Gaussian envelope, the spectral shapes of the supermodes are the successive Hermite-Gauss functions \cite{Valcarcel2012,Patera2010}. This scheme has been experimentally explored: principal modes have been determined, displaying strong squeezing on several of them \cite{Pinel2012,Roslund2015,Roslund2013} and multimode entanglement in frequency \cite{Cai2017} and time \cite{Averchenko2011}. The effective number of modes, as defined in (\ref{Nbar}), is of the order of 10. So, starting from a mode basis of single frequency modes, the number of which is roughly $10^5$, one ends up with a modal Hilbert space of a few units, a strong reduction in complexity which is useful for example when one wants to make the tomography of the generated state. Shaping the interaction by various ways like using a Spatial Light Modulator \cite{Pe2005,Patera2012,Arzani2018}, a nonlinear fiber \cite{Finger2017}, a Fabry-Perot cavity \cite{Avella2014} or optimized poling \cite{Dosseva2016} permits to modify at will the spectrum of the supermodes, and the resulting multipartite entanglement characteristics. 

\subsubsection{Above threshold operation}

 Above threshold the OPO generates bright light in the first resonant supermode (the one of largest gain, hence of largest eigenvalue $\lambda_1$). When one increases further the pump power above threshold, gain clamping prevents the other modes to oscillate \cite{Fabre2000a}, which remain with a zero mean value. It has been shown that just above threshold the supermodes different from the first one remain in squeezed vacuum states, like just below threshold \cite{Chalopin2010}. Well above threshold, the signal and idler modes carry significant energy. Pump depletion cannot be neglected and leads to the onset of a supplementary quantum coupling between the pump mode and the signal and idler modes. One has then a full three-wave mixing effect \cite{Drummond2002}, leading to three-mode entanglement between frequency modes, and six-mode entanglement between sideband modes, which have been predicted \cite{Villar2006} and observed \cite{Barbosa2018}. Note that in this regime the mode transformation is not symplectic, which implies a direct generation of a non-Gaussian state, and requires a very weak oscillation threshold of the OPO\cite{Drummond2002}.
 
Above threshold and in the case of degenerate $TEM_{01}$ and $TEM_{10}$  spatial modes, the SPOPO is predicted to generate bright light in a non-classical state \cite{Navarrete2017}, because of a symmetry-breaking effect between the transverse modes.

\subsection{Four-wave mixing}

A drawback of second-order nonlinear effects is that they are weak and exist only in centrosymmetric crystals, that can only be grown in lengths of a few centimeters. This is not the case of media with third order nonlinearities: non linear fibers can accumulate the nonlinear effect over very long lengths, and atomic media can display large effects by taking advantage of the proximity of resonances between the pump light and atomic transitions. For such media there is usually no need to insert them in a resonant cavity, and therefore the multimode quantum effects are not hampered by the mode-filtering properties of the cavity.

To produce four-wave mixing effects, one needs the simultaneous presence of two pump beams. In the simple case where one can neglect the change of the quantum state of the medium in presence of pumping light, assuming single frequency pump beams of amplitudes $\alpha_{p1}$,  $\alpha_{p2}$, and in the undepleted pump regime, the system is described by an interaction hamiltonian involving only light modes:
\begin{equation}\label{4WM}
\Hat H = \sum_{\ell, \ell'} ( A_{\ell, \ell'} \alpha_{p1} \alpha_{p2} \hat a_{\ell}^{\dagger}  \hat a_{\ell'}^{\dagger} \, + \, H. \, C.)
\end{equation}
Twin photons are therefore created in modes labeled by $\ell$ and $\ell'$. Phase matching and energy conservation requirements do not usually completely constrain the couples of twin modes, so that the light generated by four wave mixing is often multimode. The hamiltonian (\ref{4WM}) has the same structure as the parametric hamiltonian (\ref{param}), with the same consequences on the generation of twin photons and 
entangled or squeezed multimode states. 

The Kerr effect is a particular case of four-wave mixing effect. It has been one of the first techniques to generate bright squeezed beams using single transverse mode silica fibers \cite{Levenson1985}. Its multimode character in the frequency domain has been investigated  in the case of a pulsed pump \cite{Leuchs2002} and frequency combs \cite{Chembo2016}. The authors showed a nonlinear variation of squeezing with respect to spectral filtering width, which is an evidence for the generation of a multimode state, as explained in section V.C. A complete time/frequency mode analysis has been performed in \cite{Guo2015}. Here also the four-wave mixing characteristics can be appropriately tailored to meet specific purposes, for example, adjusting the spectral profile of the twin photons by using a specific pump shape and design of the photonic crystal fiber \cite{Li2012}.

In the temporal soliton regime, reached by using short and intense pulses in optical fibers, squeezing \cite{Drummond1993} and  spectral quantum correlations between different parts of the spectrum have been observed \cite{Spalter1998}, whereas in the spatial soliton regime, spatial quantum correlations between different transverse parts of the beam have been predicted \cite{Treps2000}. Such spatial correlations may  actually improve the quantum noise reduction by filtering an appropriate transverse part of the spatial soliton \cite{Mecozzi1998}.

The technique of four-wave mixing in a hot Rubidium vapor using two intense pump beams with optimized frequencies,  initiated by the NIST group \cite{Boyer2008},  provides single pass parametric gains of the order of 4 \cite{Lett2006}. It generates strong multiple correlations \cite{Wang2014}, multimode entanglement, with a record value of 9.2dB \cite{Glorieux2011} of intensity difference fluctuations below the vacuum level and localized squeezing \cite{Jing2011,Boyer2015}. It has been used for example to generated entangled images \cite{Boyer2008,Boyer2008a}. Here also, pump shaping, implemented simply by using several pump beams of different directions intersecting in the Rb cell, allow experimentalists to generate multipartite entangled beams of various topologies \cite{Qin2014,Wang2017}. The effect can be enhanced up to $9dB$ by modulating the Rb energy levels using additional lasers \cite{Zhang2017}.

Other kinds of nonlinear devices give rise to different shapes of the Schmidt modes. One can use for example 4 wave mixing in Argon-filled hollow core fibers \cite{Finger2017}.

\subsection{Multimode lasers}

Lasers generate light which can be either single mode or multimode, depending on the gain spectral profile, the properties of intermode coupling and the insertion of intracavity mode filters, but most often, their nonclassical properties are hidden by the large excess of classical fluctuations in the pumping process responsible for the population inversion. In some specific conditions, when the pump noise is greatly reduced, lasers can generate "sub-Poissonian light", i.e. beams of light having intensity fluctuations below the shot noise level. This is achieved for example by reducing the Johnson noise of the electrical current pumping high efficiency diode lasers. These diode lasers usually emit on several highly anti-correlated frequency modes \cite{Marin1995}. The same characteristics have also been observed in Vertical Cavity Surface Emitting Lasers, with are characterized by strong correlations between transverse modes \cite{Hermier1999}.

\section{Detection of multimode quantum states}

\subsection{Direct photodetection}

The observable $\hat{N}(\bm{r},t)$ associated with local photodetection, made at  point $\bm{r}$ and at time $t$ on a given beam by a photodetector of unity quantum efficiency is proportional to ${\bm \hat{E}}^{(+)}({\bm r},t)^{\dagger}{\bm \hat{E}}^{(+)}({\bm r},t)$ \cite{Glauber1963,Mollow1968}. More precisely and expressed in numbers of photon counts, using relation (\ref{decomp}), one has
\begin{equation}\label{Glauber}
\hat{N}({\bm r},t)= \sum_{n,n'}\hat{b}^{\dagger}_{n}\hat{b}_{n'}f^*_n({\bm r},t)f_{n'}({\bm r},t)
\end{equation}
Note that this observable contains crossed terms like $\hat{b}^{\dagger}_{n}\hat{b}_{n'}$. Its mean value is a linear combination of matrix elements of the coherency matrix ${\hat \Gamma}^{(1)}$. The local photodetection signal is therefore \textit{ sensitive to the correlations between the different modes}. It is only in the eigenmode basis of the coherency matrix that it appears as a sum of contributions of different modes. 

The \textit{total photodetection} observable  $\hat{N}$ is associated with the signal given by a photodetector of unit quantum efficiency  averaged over its transverse surface $S$ which is supposed to be much larger than the beam area, and over an aperture time $T$ which is supposed to be much longer than its duration. We will call such a device a "bucket detector". It is given by:
\begin{equation}\label{detection}
\hat{N}= \sum_{n} \hat{b}^{\dagger}_{n}\hat{b}_{n}
\end{equation}
thanks to the mode orthogonality when one integrates over transverse coordinates and longitudinal coordinate $z=ct$. The detector counts the total amount of photons present in the beam, which is a quantity independent of the choice of the mode basis, as we have seen above. It does not give us any information about the modal properties of the quantum state.

\subsection{Balanced homodyne detection}

\subsubsection{a mode selective detection}

One of the interests of multimode systems resides in the fact that they are likely to carry in a parallel way much more information than a single mode system. It is therefore very important to find the best way to extract from the multimode system pieces of information about any single mode of interest. Balanced homodyne detection gives precisely this possibility.

Let us consider the well-known balanced homodyne detection scheme in the context of multimode quantum optics: the beam to measure is mixed with a local oscillator (LO) on a $50\%$ beamsplitter. The observable $\hat{N}_-$, associated with the difference between the photodetection signals $\hat{N}_A$ and $\hat{N}_B$ recorded by two bucket detectors placed on output beams $A$ and $B$ and integrated over transverse space and time can be expressed as, using the well-known input-output relations of a beamsplitter:
\begin{eqnarray}
\hat{N}_-&=& \hat{N}_A-\hat{N}_B \nonumber  \\
&=& \sum_n (\hat{b}^{in \dagger}_{n,A}\hat{b}_{n,B}^{in}+\hat{b}^{in \dagger}_{n,B}\hat{b}_{n,A}^{in})
\end{eqnarray}
The latter expression is valid in any mode basis, provided the modes ${\bf f}_{n,A}({\bf r},t)$ and ${\bf f}_{n,B}({\bf r},t)$ forming the mode basis for beams $A$ and $B$ impinging on the beamsplitter are matched, meaning that the mode ${\bf f}_{n,A}({\bf r},t)$ associated with $\hat{b}_{n,A}^{in}$ is the "mirror mode" of the mode ${\bf f}_{n,B}({\bf r},t)$ associated with $\hat{b}_{n,B}^{in}$ using as a mirror the surface of the beamsplitter (symmetrical spatial shapes, identical temporal shapes). One sends on the input beam $A$ the multimode light state $\ket{\Psi}$ that one wants to characterize, and on the input beam $B$ a single mode state of light, named Local Oscillator or LO, which is made of vacuum in all modes except for a coherent state $\ket{|\alpha e^{i \phi}} $ in the $j^{th}$ mode. Let us introduce the fluctuation operators of the form $\delta \hat{O}=\hat{O} - \langle \hat{O} \rangle$. If the local oscillator intensity $|\alpha|^2$ is much larger than the vacuum fluctuations and than the mean amplitude of the multimode light beam, one can neglect all the terms except the ones proportional to $|\alpha|$. The fluctuations of the measured homodyne signal are then:
\begin{eqnarray}\label{homodyne}
\delta \hat{N}_- &\simeq&  |\alpha| (\delta \hat{b}^{in \dagger}_{j,A}  e^{i \phi}  +\delta \hat{b}_{j,A}^{in}  e^{-i \phi} )  \nonumber \\
&=& |\alpha| (\delta \hat{X}_{j} \cos \phi + \delta \hat{P}_{j} \sin \phi)= |\alpha|\,  \delta \hat{X}_{j \phi} 
\end{eqnarray}
 This equation shows that the balanced homodyne detection set-up  allows us to have access to  {\it the fluctuations of the quadrature operator $\delta \hat{X}_{j \phi}$ even if the multimode state under study has comparable or larger components on many other modes}: the homodyne detection using bucket detectors is actually \textit{a projective measurement on the LO mode}.

On the other hand, if the Local Oscillator is in a mode ${\bf g}_{LO}({\bf r},t)$ which differs from all the modes of the basis $\{{\bm f}_{n}\}$, one has, in the case of a real value of the overlap integral $\overrightarrow{f_n}^{T*}\cdot  \overrightarrow{g}_{\! LO}$  \cite{Bennink2002}:

 \begin{equation}\label{homodyne2}
\langle \delta \hat{N}_- ^2 \rangle =  |\alpha|^2 \sum_n  (\overrightarrow{f_n}^{T*}  \cdot \overrightarrow{g}_{\! LO})^2 \langle (\delta \hat{X}^{in}_{n \phi} )^2 \rangle  
\end{equation}
 \cite{Shapiro1997} have investigated the ways to optimize the LO shape in order to get the maximum squeezing effect. In \cite{Bellini2012} the single temporal mode of a heralded single photon is analyzed by homodyne detection using a pulse-shaped LO, the spectral/temporal shape of which is then algorithmically optimized in order to get the maximum overlap.

\subsection{Determination of quadrature covariance and coherency matrix elements}

We show now that it is possible, using a series of homodyne measurements with different shapes of local oscillator modes, to determine all the second order correlations functions characterizing the multimode state of light under study.

First, from the measurement of the homodyne signal variance $\langle \delta \hat{N}_-^2 \rangle$ as a function of the LO phase $\phi$, one easily extracts the quadrature variances $\langle \delta {\hat X}_n^2 \rangle$ and  $\langle \delta {\hat P}_n^2 \rangle$, as well as the XP correlation  $\langle \delta {\hat X}_n \delta {\hat P}_n \rangle$ in mode $n$. Then, in order to determine the correlations between modes $m$ and $n$, one makes another homodyne measurement, using now as the LO mode a combination of the two modes ${\bm f}_{\! LO} = ({\bm f}_{n} + {\bm f}_{m})/\sqrt2$, which gives information about the multimode state projected on this new mode, i.e. on the variances $V_{Xmn}=\langle (\delta {\hat X}_n+ \delta {\hat X}_m)^2 \rangle$ and $V_{Pmn}=\langle (\delta {\hat P}_n+ \delta {\hat P}_m)^2 )\rangle$ and on the correlation $C_{XPmn}=\langle (\delta {\hat X}_n+  \delta {\hat X}_m)(\delta {\hat P}_n+  \delta {\hat P}_m)+ (\delta {\hat P}_n+  \delta {\hat P}_m)(\delta {\hat X}_n+  \delta {\hat X}_n) \rangle/2$.

The matrix elements elements of the quadrature covariance matrix are then given by:
\begin{eqnarray}\label{serieshomodyne}
\langle \delta {\hat X}_n \delta {\hat X}_m\rangle&=& (V_{Xmn}- \langle \delta {\hat X}_n^2 \rangle -\langle \delta {\hat X}_m^2 \rangle)/2 \nonumber\\
\langle \delta {\hat P}_n \delta {\hat P}_m\rangle&=& (V_{Pmn}- \langle \delta {\hat P}_n^2 \rangle -\langle \delta {\hat P}_m^2 \rangle)/2 
\end{eqnarray}
Finally, in order to evaluate the $X_n$ $ P_m$ correlations, one uses the same procedure as the one used for the  $X_n$ $ X_m$ and $P_n$ $ P_m$ correlations, but now with another series of homodyne measurements with LO modes ${\bm f'}_{LO} = ({\bm f}_{n} +i {\bm f}_{m})/\sqrt2$ including a phase shifted $ {\bm f}_{m}$ mode.

The matrix elements of the coherency matrix ${\bf \Gamma}^{(1)}$ can also be determined from the series of homodyne measurements  using relations (\ref{serieshomodyne}) and (\ref{compcov3}).

It is thereby possible to get the full correlation matrices by a series of homodyne measurements (for example one must use 100 choices of different LO modes for 10 modes). These measurements must be made sequentially, and require therefore that the quantum state generator is stable over the duration of these measurements. The quantum state which is characterized by this  matrix is unfortunately destroyed by the measurement, which prevents this technique to be used in experiments involving conditional measurements, and in particular in Measurement Based Quantum Computing.

 Note that, with the help of a beamsplitter and two homodyne detectors at its two outputs, it is possible to measure at the same time $\langle \delta X_j^2 \rangle$ and $\langle \delta P_j^2 \rangle$, but this cannot be done without adding excess noise coming from vacuum fluctuations at the input of the beamsplitter, as $X_j$ and $P_j$ are non-commuting quantities that cannot be measured exactly simultaneously. This method is useful to measure classical noises that are significantly larger than the vacuum fluctuations. One can then extract from these measurements the principal noise modes that govern the dynamics of a classical light source. This has been in particular achieved for mode-locked lasers \cite{Schmeissner2014}. 

\subsection{Spectral homodyne and resonator detection}

In the case of stationary light sources (continuous wave or periodic), homodyne detection is very often followed by a spectral analysis of the fluctuating signal, which allows us to measure the noise spectral density $S(\Omega)$ of the homodyne signal fluctuations. For example when the LO phase $\phi$ is zero,  $S(\Omega)$  is equal to $S(\Omega)=\langle \delta {\hat X}_{\Omega}^2 \rangle $ where 
\begin{equation}
\delta {\hat X}_{\Omega}= \frac1{\sqrt{2 \pi}}\int \, dt \, e^{i\Omega t} \delta {\hat X} (t)
\end{equation}
It is easy to show that this measured signal depends on the properties of \textit{sideband modes}, which are quasi-monochromatic frequency modes at frequency  $\omega_0 \pm \Omega$ (within the frequency bandwidth $\Delta \Omega$ of the spectrum analyzer), where  $\omega_0$ is the optical carrier frequency and  $\Omega$ the Fourier analysis frequency, usually in the MHz range. More precisely, if one calls ${\hat a}_{\omega}$ the annihilation operator in the frequency mode of frequency $\omega$, one has obviously:
\begin{equation}
\delta {\hat X}_{\Omega}={\hat a}_{\omega_0+ \Omega} + {\hat a}_{\omega_0 - \Omega}
\end{equation}
A noise spectrum of the homodyne signal, apparently a technique to characterize single mode fields, actually gives a highly multimode information about couples of sideband modes. For example squeezing of noise frequency components in a given Fourier frequency range ($\langle \delta {\hat X}_{\Omega}^2 \rangle <1$ for $\Omega \in [\Omega_1, \Omega_2]$) can actually be seen as $(\Omega_2 - \Omega_1)/\Delta \Omega$ independent couples of EPR correlated sideband modes. 

However, it can be shown that spectral homodyne detection does not give the full information about the sideband modes \cite{Barbosa2013}, more precisely it measures the properties of the input quantum state partially traced over the mode associated with ${\hat a}_{\omega_0+ \Omega} -  {\hat a}_{\omega_0 - \Omega}$, orthogonal to $\delta {\hat X}_{\Omega}$. There is another detection technique, named "resonator detection " which gives indeed access to the full information about the frequency modes. It consists in measuring the intensity fluctuations of the beam to analyze after it has been reflected on a slightly off-resonant Fabry-Perot cavity the length of which is scanned. The off-resonance unbalances the two sideband and reveals a possible asymmetry between them, which is not possible with the homodyne detection. Let us mention that sideband modes, although very close to each other in frequency, can be separated and studied individually using interferometric techniques \cite{Huntington2005}

\subsection{Multiplexed detection}

\subsubsection{Implementation}

Homodyne detection is a destructive measurement, which prevents any further processing of the same quantum state, so a "single shot" detection of several observables on the same quantum system, that we will call "multiplexed detection", is highly desirable, especially if one wants to take advantage of the multimode aspect of the generated light. A multiplexed detection can be simply implemented when the different modes of the light can be spatially separated: one inserts homodyne detection devices on all, or some, of these different modes \cite{Su2014,Yukawa2008,Su2012}. In this configuration the different measurements do not completely destroy the quantum state. One can also post select a subset of recorded data in order to herald a specific quantum state \cite{Aichele2002, Laurat2003}. One can also use the information contained in the measurements to correct by a feedforward technique a mode which has been left un-measured, for example in sub-Poissonian bright beam generation \cite{Mertz1990} or in teleportation  \cite{Furusawa1998} .

When the multimode state is propagating in a single spatial beam, different kinds of multiplexed photodetectors can be used:  CCD cameras \cite{Perina2012} or photodiode arrays \cite{Beck2000,Dawes2001,Armstrong2012} for spatial modes, and time resolved detectors for temporal modes. They allow to record the intensity fluctuations on pixels of area $\delta x^2$, time bins of duration $\delta t$ and the correlations between the fluctuations of different pixels. In addition, if one inserts a dispersive device like a prism in front of the array detector, one can measure the fluctuations of different frequency bands of spectral width $\delta \omega$ \cite{Ferrini2013}.

\subsubsection{Multiplexed homodyne signal}

The modes corresponding to the detection scheme we just described are respectively the pixels, time bins and/ frequency band modes: they are normalized modes ${\bf v}_{j}({\bf r},t)$ which are zero outside the detection domain (in space, time or frequency) and are constant inside, that we will call in a general way  "bin-modes". Let us consider the set $\left({\bf v}_{j}({\bf r},t)\right)$ of such bin modes which are not overlapping and which cover the whole detection domain, and the corresponding annihilation operators ${\hat d}_j$. Such a detection acts actually as a low pass filter: the  bin-modes constitute a complete  and orthonormal set of modes in the subspace of electric fields which have an upper limit $1/2 \delta x$ in the case of pixels, or $1/2 \delta t$ in the case of time bins, or $1/2 \delta \omega$ in the case of frequency bands. The intensity $\hat{N}_j$, in terms of photon number, recorded on the bin-mode labeled by $j$, deduced from (\ref{Glauber}), is:
\begin{equation}\label{detectionpixel}
\hat{N}_j=  \hat{d}^{\dagger}_{j}\hat{d}_{j}
\end{equation}
because the other modes vanish on the $j^{th}$ detector. This relation, without  summation over modes, is only valid for operators  $\hat{d}^{\dagger}_{j}\hat{d}_{j}$ defined in the bin-mode basis.

Let us now consider the balanced homodyne detection set-up that uses a single mode LO in quantum state $|\Psi_{LO}\rangle$ and multiplexed detectors instead of bucket detectors. The measurement of the intensity difference between analog pixel/time/frequency bins $j$ on the two output beams of the beam splitter is now given by, omitting for simplicity the superscript $in$:
\begin{equation}\label{homodynemultiplex}
\delta \hat{N}_{-,j} \simeq \langle \Psi_{LO} |\hat{d}_{j,b} |\Psi_{LO}\rangle \delta \hat{b}^{\dagger}_{j,a}  + h.c.  \end{equation}

We take as before the LO to be a coherent state $| |\alpha| e^{i \phi} \rangle$ in mode $\overrightarrow{g}_{\! LO}$, assumed to vary slowly over the extension of a bin. We have seen that  it may be also written as a tensor product of coherent states in any other mode basis, including the bin basis. One can show:
\begin{equation}\label{LO2}
\langle \Psi_{LO} |\hat{d}_{j,b} |\Psi_{LO}\rangle=(\overrightarrow{v_j}^{T*} \cdot \overrightarrow{g}_{\! LO})  |\alpha|  e^{i \phi}
 \end{equation}
 Assuming the modal inner product to be real, one has finally:
\begin{equation}\label{homodynemultiplex2}
\delta \hat{N}_{-,j} \simeq  |\alpha|  (\overrightarrow{v_j}^{T*} \cdot \overrightarrow{g}_{\! LO}) \delta \hat{X}_{j \phi}
\end{equation}
The condition of validity of this expression is that $ |\alpha|  (\overrightarrow{v_j}^{T*} \cdot  \overrightarrow{g}_{\! LO})$ is large enough compared to vacuum fluctuations. Its exact value is taken care of by a proper normalization of the homodyne signal.

A computer memory then stores in parallel the instantaneous different detection signals  $\delta \hat{N}_{-,j}$ for all the bin modes and for a given LO phase $\phi$, for example $\phi=0$, which give access to the ${\hat X}_j$ quadrature. By a time integration of the square of the fluctuations over a time window $T$, one then gets in a single shot all the variances  $\langle \delta X_j^2 \rangle$; by a time integration of the product of the recorded fluctuations $\delta X_j$ and  $\delta X_{j'}$ one gets the cross correlations $\langle \delta X_j  \delta X_{j'} \rangle$ i.e. a quarter of the quadrature covariance matrix. The second moments of the $P$-quadrature are obtained by a second single shot measurement with a LO phase $\phi=\pi/2$. In order to measure the cross-correlations $\langle \delta X_j  \delta P_{j'} \rangle$ for all $j'$, one needs to dephase by $\pi/2$ the part of the Local Oscillator which will impinge on bin $j$ after the beamsplitter. This can be done using a Spatial Light Modulator (SLM) or a phase plate on the LO beam. . 

If the LO phase $\phi$ is varied on a time scale shorter than $T$, one gets phase averaged correlations, which give directly the value of  $\langle \delta X_j  \delta X_{j'} +\langle \delta P_j  \delta P_{j'} \rangle$, i.e.  the real part of the coherency matrix.

\subsubsection{Multiplexed mode discrimination via post processing}

Let us consider now more generally a multimode quantum state $|\Psi \rangle$, described in a given mode basis $\{ {\bm u}_l \}$, that one wants to characterize through multiplexed homodyne detection. The question is which set of quadrature operators, and for which mode basis, can be accessed simultaneously for a given measurement scenario. A multipixel homodyne detection performs a measurement in the bin mode basis $ \{ {\bm v}_j \}$ which is related to the initial mode basis through a modal unitary operator $U_b$. This matrix depends on the optical arrangement between the quantum state to be characterized and the detector. It can be adjusted by changing the experimental setup. But for a given setup, shaping the local oscillator and post-processing on a computer allows for a large variety of quadrature outcomes. In particular, one can:

\begin{itemize}

\item shape the local oscillator impinging on each pixel of the multipixel detector, and thus the modal projection. This induces a change in the bin mode basis $ \{ {\bm v}_j \}$ which is measured. This effect can be mathematically included in the matrix $U_b$.

\item add a phase shift $e^{i \psi_j}$ on each bin mode ${\bm v}_{j}$. This is done phase-shifting by $e^{i \psi_j}$ the local oscillator beam impinging on the corresponding pixel of the multipixel detector. We call $\Delta_{LO}$ the associated diagonal unitary operation on the modes;

\item digitally recombine the electronic signals coming from each bin by multiplying them with real gains, amounting to applying an orthogonal matrix $O$ to the vector of measured quadratures of modes  $\{ {\bm v}_{j} \}$.

\end{itemize}

Let us define the mode basis:
\begin{equation}
\overrightarrow{c}_n =\sum_{l=1}^{N_p} (O \Delta_{LO} U_b)_{n}^{l} \overrightarrow{u}_l= \sum_{l=1}^{N_p} (U_{MPHD})_{n}^{l} \overrightarrow{u}_l
\end{equation}
where $Np$ is the number of pixel detectors. The data processing technique that we just described allows us to access simultaneously the amplitude quadrature fluctuations in the set of modes $\{ {\bm c}_n \}$, \textit{even though such modes have not been physically extracted from the multimode beam} \cite{Ferrini2013}. This technique is very useful in order to have access to the nodes of cluster states which are embedded in the multimode quantum state (see section XIV). It can be shown \cite{Armstrong2012} that many, but not all, possible modal unitary transformations $U$ can be emulated by this procedure, namely those such that $U_bU^T UU_b^\dagger=D$, where $D$ is a diagonal matrix with unit modulus complex elements \cite{Ferrini2013}. 

Finally, let us stress that data processing is not limited to linear combinations of data. Non-linear ones can be also performed, in order to implement non-Gaussian operations by feedforward techniques \cite{Miyata2016}.

\subsection{Two-photon detection}

In addition to quantum fluctuations measurements, joint two-photon detection is a privileged tool to reveal the non-classical properties of a quantum state of light. The observable $\hat{W}(\bm{r},t, \bm{r'}, t')$ associated with the double detection at times $t$ and $t'$ made by two detectors A and B  at  positions $(\bm{r}, t)$ and $(\bm{r'}, t')$ is \cite{Glauber1963}:
\begin{equation}\label{Glauber2}
\hat{W}(\bm{r},t, \bm{r'}, t')= \hat{E}^{(+) \dagger}({\bm r},t) \hat{E}^{(+) \dagger}({\bm r'},t') \hat{E}^{(+)}({\bm r'},t') \hat{E}^{(+)}({\bm r},t) 
\end{equation}

When integrated over time and transverse space, it gives a "total double click" observable equal to:
\begin{equation}\label{Glauber3}
\hat{W}=\sum_{m_A, m_B}{\hat b}^{A \dagger}_{m_A}{\hat b}^{B \dagger}_{m_B}  {\hat b}^A_{m_A}{\hat b}^B_{m_B}
\end{equation}
which does not depend on the choice of the mode basis and has a mean value which is zero for a single photon state, and equal to the square of the mean photon number for any multimode coherent state (as defined in section V.A c). 

Let us now consider the Hong-Ou-Mandel (HOM) configuration \cite{Hong1987}: the two detectors are measuring photon coincidences between the two outputs of a $50\%$ beamsplitter. More details concerning the derivations can be found in appendix C. We assume in addition that the detectors are "slow", so that they lose the information about the exact arrival time of each photon. We can then use formula (\ref{Glauber2}) for a "bucket detector" obtained by integration over times $t$ and $t'$.

$\quad$ a) Let us first consider the case where two uncorrelated single photons in modes $g_A$ and $g_B$ are impinging on the two input ports of the beamsplitter \cite{Bylander2003,Legero2003,Beugnon2006,Kaltenbaek2006}. The  calculation, outlined in Appendix C, yields the following expression for the normalized two-photon detection rate for zero difference between the two photon paths, $g^{(2)}(0)$:
\begin{eqnarray} \label{g2}
g^{(2)}(0)&=&\frac12 \left( 1 -  |\overrightarrow{g_A}^{T*} \cdot \overrightarrow{g_B} |^2 \right) \nonumber\\
&=&\frac12 \left( 1 -  |\langle 1: {\bm g}_A | 1: {\bm g}_B \rangle |^2 \right)
\end{eqnarray}
One observes that the HOM destructive interference is perfect only when $|\overrightarrow{g_A}^{T*} \cdot \overrightarrow{g_B} |^2 =1$, implying that  $\overrightarrow{g}_A= \overrightarrow{g}_B$ within a phase term \cite{Ou2017,Beugnon2006}: single photons "coalesce" on the beamsplitter only when they are in modes which are strictly identical, both for their space and time dependences. If the modes are orthogonal, one retrieves the classical value $1/2$ for the normalized coincidence rate.

$\quad$ b) There is a second physical situation for HOM interference, which was actually the one of the initial experiment \cite{Hong1987}, consisting in using entangled two-photon states generated by parametric down conversion. The calculation is outlined in Appendix C. Using the Schmidt decomposition of the input state one finds that the coincidence rate for zero path length difference between the two arms vanishes not only when the input quantum state is a product of single mode states, but also when it is entangled, provided all the Schmidt  modes of the parties $A$ and $B$ have identical space-time dependences: $\forall i \, \overrightarrow{g}_A^{i} =\overrightarrow{g}_B^{i}$. Such a perfect two-by-two matching for all the Schmidt modes of the  two beams is achieved when there is total symmetry with respect to the exchange between the signal and idler A and B parts, i.e. when the matrix $G_{si}$ introduced in (\ref{Gsignalidler}) is symmetric.

\section{Multimode amplification and attenuation}

\subsection{effect on squeezing}

The quantum aspects  of single-mode attenuation and amplification have been studied and understood for a long time \cite{Caves1982,Caves2012}: losses, as well as phase insensitive amplification, lead to a necessary coupling between the considered mode and the outer world, which results in added noise, meaning reduction of squeezing and a minimum 3 dB noise penalty in the noise figure of the amplifier. Only phase-sensitive amplifiers are authorized to be noiseless by the laws of Quantum Mechanics \cite{Marino2012}. It is therefore important to see in which respect these properties extend to the multimode case \cite{Lane1983}, considering the fact that many kinds of multimode optical amplifiers have been developed in the recent years, for example Erbium-doped fibers amplifiers \cite{Nykolak1991}, parametric amplification in crystals \cite{Allevi2006} or in fibers \cite{Guasoni2016},  image amplifiers based on four wave mixing in atomic vapors \cite{Boyer2008a, Gigan2005,Ferrini2014}. 

In this section, we follow mainly the argument given in \cite{Leuchs2006}.  We restrict ourselves to the case of a phase insensitive multimode attenuator or amplifier, which is characterized by an intensity multiplicative factor $P$, acting on a N-mode modal space having  mode basis $\overrightarrow{f}_m$ and corresponding annihilation operators ${\hat b}_m$. At the classical level, the output field $E^{(+) out}$ is equal to  $\sqrt P \, E^{(+) in}$ for any input $E^{(+) in}$. At the quantum level, the corresponding relation for the column vector of annihilation operators in the Heisenberg representation 
\begin{equation}
\overrightarrow{{\hat b}}^{out}=\sqrt P \,\overrightarrow{{\hat b}} ^{in}
\end{equation}
cannot be valid, except when $P=1$, because it is not a commutator preserving relation. One therefore needs to introduce a set of $N'$ ancilla modes,  characterized by a mode basis $\overrightarrow{g}_n$ and corresponding annihilation operators ${\hat a}_n$, which are coupled to the amplifier modes. The actual input-output relation writes then:
\begin{equation}\label{inout}
\overrightarrow{{\hat b}}^{out}=\sqrt P \, \overrightarrow{{\hat b}}^{in} + \mathcal{L}\overrightarrow{{\hat a}}^{in} + \mathcal{M} \overrightarrow{{\hat a}}^{in, \dagger}
\end{equation}
where $\mathcal{L}$ and $\mathcal{M}$ are $N'$ lines $N$ columns matrices. The canonical commutation relations $[\overrightarrow{{\hat b}}^{out},\overrightarrow{{\hat b}}^{\dagger out}]= {\bf 1}_N$ are ensured when:
\begin{equation}\label{noisemode} 
\mathcal{M}\mathcal{M}^{\dagger}- \mathcal{L}\mathcal{L}^{\dagger} = ( P-1 ) {\bf 1}_N 
\end{equation}
Let us now consider the amplifier case $P>1$ and define a new column vector of operators:
\begin{equation}\label{noisemode1}
\overrightarrow{{\hat c}}^{in}=\frac1{\sqrt {P-1}} \, (  \mathcal{L}^{\dagger} \overrightarrow{{\hat a}}^{in,\dagger}+ \mathcal{M}^{\dagger} \overrightarrow{{\hat a}}^{in})
\end{equation}
One deduces from relation (\ref{noisemode}) that  $[\overrightarrow{{\hat c}}^{in},\overrightarrow{{\hat c}}^{in,\dagger}]= {\bf 1}_N$, so that $\overrightarrow{{\hat c}}^{in}$ is a column  vector of bosonic annihilation operators, associated with modes that can be called noise modes. One can finally write:
\begin{equation} \label{out}
\overrightarrow{{\hat b}}^{out}=\sqrt P \, \overrightarrow{{\hat b}}^{in} + \sqrt{P-1} \overrightarrow{{\hat c}}^{in, \dagger}
\end{equation}
By definition $\overrightarrow{{\hat b}}^{out}$ and $\overrightarrow{{\hat b}}^{in}$ are of dimension $N$, whereas $\overrightarrow{{\hat c}}^{in}$ is of dimension $N'$, which implies that $N'=N$.

In the attenuator case $P<1$ one has similarly:
\begin{equation} \label{noisemode2}
\overrightarrow{{\hat b}}^{out}=\sqrt P \, \overrightarrow{{\hat b}}^{in} + \sqrt{1-P} \,  \overrightarrow{{\hat d}}^{in}
\end{equation}
with the operators $\overrightarrow{{\hat d}}^{in}=\frac1{\sqrt {1-P}} \, ( \mathcal{L}\overrightarrow{{\hat a}}^{in}+  \mathcal{M} \overrightarrow{{\hat a}}^{in,\dagger})$ satisfying bosonic commutation relations  $[\overrightarrow{{\hat d}}^{in},\overrightarrow{{\hat d}}^{in,\dagger}]= {\bf 1}_N$ and $[\overrightarrow{{\hat d}}^{in},\overrightarrow{{\hat d}}^{in}]=0$. 

We have therefore found that in the general case of $N$-mode linear optical systems written in any mode basis,  there are $N$ associated ancilla modes which independently bring excess noise to the "useful" modes. These noise modes depend on the physical system considered and are not necessarily associated with optical modes.

If the noise modes are in the vacuum state, one can write a simple relation for the evolution of the coherency matrix, as well as for the quadrature covariance matrix valid in any mode basis and in both the attenuator and amplifier cases:
\begin{eqnarray} \label{amplicov}
{\bf {\Gamma}}^{(1), out} &=& P \, {\bf {\Gamma}}^{(1), in} + |P-1|  {\bf 1}_N \\ {\bf {\Gamma}}_Q^{out} &=& P \, {\bf {\Gamma}}_Q^{in} +  |P-1|   {\bf 1}_{2N} 
\end{eqnarray}
As in the single mode case, an energy gain P of 2 is enough to bring above shot noise a perfectly squeezed input state, even multimode. 

Equation (\ref{amplicov}) is not valid for a phase sensitive amplification, for which the 3dB penalty does not exist.   Amplification with reduced added noise has been observed in the case of parametric amplification of optical images \cite{Mosset2005}

\subsection{effect on entanglement}

We can now determine whether entanglement is preserved or not under such a linear processing. For bipartite entanglement between modes 1 and 2, we may use the Duan-Mancini criterion: 
\begin{equation}
\langle \left(\frac{{\hat X}_1 + {\hat X}_2}{\sqrt2}\right)^2\rangle \langle \left(\frac{{\hat P}_1 - {\hat P}_2}{\sqrt2}\right)^2\rangle <1 
\end{equation}
For a maximally entangled input state, then the l.h.s. quantity, which is zero at the input of the amplifier, is equal, according to (\ref{amplicov}) to $(P-1)^2$  at its output: entanglement survives for any attenuation factor $P<1$, with a decreasing violation of the inequality when the losses increase.  In addition, like for perfect squeezing, entanglement survives for gains smaller than 2. When $P \ge 2$ then the criterion is no longer satisfied. If one assumes that there are no X-P correlations, then the Duan criterion is necessary and sufficient for Gaussian states, and one is sure that entanglement "dies" for gains higher than 2. Squeezing and entanglement are equally destroyed by amplification and attenuation. This is another proof that they are indeed the two faces of the same physical property , which has two different "avatars" according to the choice of the mode basis. 

\section{Mode shape control}

An important property of optical modes is their ability to be shaped at will so as to match as well as possible the spatio-temporal dependence needed in a given application, such as optimized parameter estimation (section XIII), coupling with quantum memories or reconfigurable quantum information processing (section XIV). 

Mode shaping can be implemented in different ways:

\begin{itemize} 
\item as a mode converter, which transforms a given single mode input to another single mode output, while keeping unchanged the quantum state defined in the mode ;

\item as a mode extractor, which filters a mode of interest from a multimode field, while keeping its quantum properties and its correlations with the not-extracted modes. Such a device would enable us to keep this mode for further use, for example to correct it in a feedforward scheme. This is easy to do of course when the modes are spatially, spectrally or temporally separated, with a mutual "distance" large enough to be separable by the current technology, but not when the mode are overlapped;

\item  as a mode multiplexer/demultiplexer which converts in a parallel way a set of input modes into an orthonormal basis of modes that are easy to propagate in a given system, and makes the reverse process at its output, while keeping the intrinsic properties of the state, and in particular the principal modes and their squeezing performances. As an example in classical optics, in Wavelength Division Multiplexing the different input frequency channels are merged in a single mode fiber (multiplexing) and are physically separated at its output (demultiplexing) using dispersive devices. An even more advanced concept is a programmable network in which this whole process can be modified at will according to the chosen application.
  
\end{itemize} 

This domain of optics is very active, especially with the recent availability of Spatial Light Modulators (SLM), and both at the classical and quantum levels. We will of course focus on its quantum aspects, and will be concerned by the evolution of the modes, but also by the induced evolution of the quantum state of the system when the modes in which is it defined are converted or distorted by some optical process.

For quantum applications, in the ideal case, the mode shaping must be lossless.  Let us call $p$ a parameter governing the change of the boundary conditions inducing the mode shaping. If the mode transformation conserves energy  it cannot change the frequency of the photon. The energy levels of the quantum optical system are then independent of $p$ and do not get close to each other when $p$ is changed. If $p$ is swept slowly compared to the optical period, we can safely assume that the state transformation induced by the mode transformation is adiabatic and therefore that the quantum state does not change during the mode conversion. In addition, if the mode conversion is passive and linear, one can reasonably assume that it conserves the number of photons : an input $n$-photon Fock state $| n: f_{in} \rangle $ must be then transformed into the same Fock state in the output mode $| n: g_{out} \rangle $ whatever $n$. This is therefore true for any input state, because Fock states form a basis of the Hilbert space of quantum states.

In the following we will separately describe mode shaping effect effected by linear optics and by nonlinear optics. 

\subsection{Linear conversion}

Let us first consider spatial modes: it has been mathematically proved in the general case \cite{Morizur2010} that one can convert any input spatial mode to any output spatial by using reflections (or transmissions) on two appropriately chosen phase plates separated by free space. One needs a larger, but finite, amount of reflexions on successive phase plates in order to convert a whole spatial mode basis into another arbitrary mode basis. This can be implemented with low conversion losses \cite{Labroille2014} in a device called "Multiplane Light Converter" (MPLC) with the help of SLMs or phase plates. There are of course other linear optical spatial mode converters, but they are useful only for a subset of transformations: lens and free propagation for spatial Fourier transform,  telescope for beam magnification, Babinet-Soleil-Bravais birefringent filter for spatial derivative \cite{Labroille2013}. Fractional Fourier transform allows to design Laguerre-Gauss mode sorters \cite{Zhou2017}, which, combined with an astigmatic mode converter from Hermite-Gauss to Laguerre-Gauss modes consisting of cylindrical lenses \cite{Allen1993},  can also sort Hermite-Gauss modes \cite{Zhou2018a}. Tapered fibers down to nanometer scale \cite{Tong2003}, or photonic lanterns  \cite{Fontaine2012} are also used to manipulate modes of fibers. A SLM can also be used to "preform" the mode before it interacts with a random scattering medium in order to obtain at its output a well focussed beam having kept its quantum properties \cite{Defienne2016}. 

If one now considers time/frequency modes, passive unitary conversion is not possible from a given input frequency mode to a frequency shifted  output mode, because changing the frequency of light amounts to changing the photon energy. This cannot be done by a passive  transformation, whereas in the spatial case, changing the direction of propagation of a photon does not change its energy. So passive unitary frequency mode conversion can only involve phase changes at the different frequencies, which are implemented using SLMs inserted between a pair of diffraction gratings and lenses \cite{Weiner2011}. Such linear devices are also used in a non-unitary way by implementing frequency dependent losses to shape the light spectrum. These last devices destroy the quantum properties of the quantum state in the CV regime as shown in section XI, and the probability of single photon counts in the DV regime. In addition, they will never be able to induce a broadening of the spectrum. This kind of lossy filtering technique allows for multiplexing/demultiplexing techniques, for example for single photons in the multiple spatial modes of a silicon photonic crystal fiber \cite{Carpenter2013}, or temporal \cite{Perez2015}. Long dispersive fibers provide an efficient way to make a wavelength to time mapping of a multimode input light  pulse \cite{Chandrasekharan2017}.

Mode extractors are also very important tools to handle modes in view of applications: it is for example possible to filter a given Hermite-Gauss mode by using a Fabry-Perot cavity which transmits one mode and reflects all the other ones. One also extracts Laguerre-Gauss modes using especially designed phase holograms \cite{Ren2017}. The MPLC device also allows to multiplex/demultiplex a number of orthogonal spatial modes, of the order of 10.

\subsection{Non-linear conversion}

If one wants to include frequency changes in the mode transformation one must rely on nonlinear effects, either in the microwave domain or in the optical domain. Let us consider as an example the sum-frequency generation (SFG) \cite{Eckstein2011}, which uses a $\chi^{(2)}$ type II non linear medium (equation \ref{param}). It is here pumped by a low frequency gate beam of spectral amplitude $\alpha_g (\omega)$, treated classically. The corresponding Hamiltonian is

\begin{equation}\label{SFG}
\Hat H' =\sum_{\ell_1, \ell_2} ( G'_{\ell_1, \ell_2} \hat a_{\ell_1} \hat a_{\ell_2}^{\prime \dagger} \, + \, H. \, C.)
\end{equation}
where $\hat a_{\ell_1}$ is the annihilation operator of input photons in  the single frequency mode ${\bm f}_{\ell_1}$ of frequency $\omega_{\ell_1}$,  and $ \hat a^{\prime}_{\ell_2}$ is the annihilation operator of SHG photons in  the single frequency mode ${\bm f'}_{\ell_2}$ of frequency $\omega_{\ell_2}$.  $G'_{\ell_1, \ell_2}= a'_{\ell_1, \ell_2} \alpha_g(\omega_{\ell_2} - \omega_{\ell_1})$. This Hamiltonian is a double sum of terms describing transfer processes from mode ${\bm f}_{\ell_1}$ to mode ${\bm f'}_{\ell_2}$ operating in very different spectral ranges. A Singular Value Decomposition (SVD) of matrix $G'$ leads to the following expression of the Hamiltonian, analog to the one describing a beamsplitter:
\begin{equation}\label{SFG2}
\Hat H' =\sum_i ( \mu_i  \hat b_{i} \hat b_{i}^{\prime \dagger} \, + \, H. \, C.)
\end{equation}
$\hat b_{i}$ and  $\hat b'_{i}$ are respectively annihilation operators of photons in  Schmidt input mode ${\bm g}_{i}$ and "twin" Schmidt mode ${\bm g'}_{i}$. The spectral shape of these Schmidt modes depends on the matrix $G'_{\ell_1, \ell_2}$ and therefore on the pump, or gate, spectral shape  $\alpha_g (\omega)$ and on the phase matching properties of the nonlinear crystal. If the gate has a Gaussian shape $HG_0$, the Schmidt modes ${\bm g}_{i}$ and ${\bm g}^{\prime}_{i}$ are both Hermite-Gauss spectral/temporal  modes of same index $HG_i$  \cite{Eckstein2011,Ansari2018}. If the group velocities of the input and gate modes are matched, the Singular Value Decomposition leads to a single non negligible coefficient in the Schmidt sum.  More precisely if the pump is in the mode $HG_j$ , then the single input Schmidt mode is also $HG_j$, whereas the SFG Schmidt mode is the Gaussian mode $HG_0$ whatever the pump.

In the last configuration, if the gate is weak, and when there is only one term, of index $1$, in the sum (\ref{SFG2}) the evolution operator writes:
\begin{equation}\label{SFG3}
\Hat U =\hat 1 - i  \mu'_1  (\hat b_{1} \hat b_{1}^{\prime \dagger} \, + \, \hat b_{1} ^{\dagger} \hat b_{1}^{\prime})
\end{equation}
where $\mu'_1 = \mu_1  L/\hbar c$, $L$ being the crystal length. If the input state is for example a single photon state ${\bm g}_{1}$, $ |1: {\bm g}_{1}\rangle$, the output state $|\Psi_{out}\rangle$ is:
\begin{equation} \label{SFG1}
|\Psi_{out}\rangle= |0 : {\bm g}^{\prime}_1 , 1: {\bm g}_{1} \rangle - i \mu'_{1}  |1: {\bm g}^{\prime}_{1}, 0: {\bm g}_{1} \rangle
\end{equation}
where $\mu'_i = \mu_i  L/\hbar c$, $L$ being the crystal length: the single photon has been transferred from Schmidt mode $ {\bm g}_{1}$ to the Schmidt twin mode mode at the double frequency ${\bm g}^{\prime}_{1}$. 
If one increases the gate pump beam power, but still in the undepleted pump approximation, the input output relations for the annihilation operators of the input and SHG Schmidt modes are:
\begin{eqnarray}\label{SFG4}
 \hat b_{1}^{out} &=& i \sin \mu'_1 \, \hat b_{1}^{\prime in} + \cos \mu'_1   \, \hat b_{1}^{in} \nonumber \\
  \hat b_{1}^{\prime out} &=& \cos \mu'_1  \, \hat b_{1}^{\prime in} + i \sin \mu'_1 \, \hat b_{1}^{in}
\end{eqnarray}
 When $\mu'_i$ is equal to $\pi/2$, then $ \hat b_{1}^{\prime out}= i \hat b_{1}^{in}$. We have here a \textit{perfect mode extractor}: only one input mode $ {\bm g}_{1}$ is perfectly transferred to SHG mode  ${\bm g}^{\prime}_{1}$, whereas all the other modes are left unchanged by the nonlinear process. One easily changes the mode which is extracted by changing the shape of the pump gate beam.  One can therefore extract selectively from a multimode  input beam a given Hermite-Gauss mode $HG_j$ by using a gate beam precisely in the mode $HG_j$. This device is often called a "Quantum Pulse Gate" and works whatever the quantum state which "dwells" in the input Schmidt mode. 

So Sum Frequency Generation through three- and four-wave mixing is an ideal tool to change the mode of a given quantum state  \cite{McKinstrie2012} . It has been mainly implemented for single photon states and weak coherent states, for example to shift them from a wavelength range to another with good conversion efficiency \cite{Tanzilli2005}, to select a given mode (spatial or temporal) with efficiencies of the order of $70\%$ \cite{Brecht2014,Reddy2018} and good fidelity for the output state \cite{Guinness2010}, or to manipulate specific  temporal modes among many others \cite{Ra2017,Perez2015,Reddy2014}. Let us mention that other nonlinear processes, such as frequency down conversion, can be used to transfer single photon states \cite{Curtz2010,Lenhard2017} in an entanglement preserving way. Cavity QED effects can also be used to shape at will temporal modes of single photons \cite{Morin2019}.

Temporal focussing and imaging of a non-classical state is an interesting issue  \cite{Kolner1989}. It can be implemented  also  in a noiseless way using sum frequency generation \cite{Patera2018}, so that squeezing can be preserved by the operation, with a change of the Fourier spectrum of the quantum fluctuations. One can also use electro-optic modulation as a unitary time lens able to compress the spectral width, and enhance the peak intensity of single photon states \cite{Karpinski2017}.

\section{Mode optimization in parameter estimation}

Light is often used as a tool to perform very accurate or sensitive measurements of some parameter, like distance, velocity, time delay, frequency.... We will name $a$ this parameter in a generic way. It is important to know what is "the best light" which will enable us to make the best estimation of $a$. We have of course the choice of the quantum state of the light, but also of the spatial and temporal shape of the mode(s) in which this state is defined. So far, most attention has been given to the quantum state issue, and there is an extensive literature on its choice \cite{Giovannetti2011,Wiseman2009}, which constitutes an important part of the domain of quantum metrology, but much less attention has been given to the mode issue.
 
To estimate a parameter $a$ by optical means, one needs first an optical system that generates a $a$-dependent beam of light. This light is measured by one way or another, yielding data that are then processed in order to derive an estimator $\tilde a$ of $a$, from which a value of $a$ is inferred. Generally speaking, the light used in the measurement may be multimode, the detection may be multiplexed, and the data processing may involve the analog or numerical processing  of the measured quantities. We will restrict ourselves to non-biased estimators, in which case the quality of the measurement is evaluated by its "sensitivity", i.e. by the standard deviation $\Delta a$ of the estimated values of $a$ around the "true" value of the parameter (that we will take for simplicity to be $0$), which gives an upper limit to the smallest measurable variation of $a$.

We will concentrate on pure states, less "noisy" than mixed states, and call $| \Psi(a)\rangle$ the (possibly multimode) quantum state of the light that is submitted to measurement. The Quantum Cram\'er-Rao limit \cite{Helstrom1998,Helstrom1967,Helstrom1969,Braunstein1994} allows us to find the smallest value of the standard deviation $\Delta a$ of the estimated values optimized over all possible processing procedures of the experimentally recorded data and over all possible optical measurements performed on the {\it a}-dependent beam. But there is so far no known optimization procedure over all possible quantum states of light, and many different authors have proposed many possible non-classical quantum states giving a "quantum advantage" in parameter estimation. Note that the mode optimization issue in parameter estimation is related to the problem of state discrimination  \cite{Pirandola2008}. 

We will here restrict ourselves to the subset of \textit{multimode Gaussian states}\cite{Pinel2012,Pinel2012a,Pinel2013,Safranek2015,Nichols2018} (section VII). Such a choice excludes highly non-classical states like Fock or NOON states, but includes squeezed and EPR entangled states. It has the interest of comprising also the bright coherent states which can be readily experimentally produced with mean photon numbers $N$ as high as $10^{15}$, which is far from being the case for Fock or NOON states. Two  modes $u_{mean}$ and $u_{det}$ play an important role in the present problem: 
\begin{eqnarray}\label{mean}
u_{mean}({\bf r}, t, a) &=& \frac{1}{\mathcal{E}^{(1)}\sqrt{N}}  \langle \Psi(a) | {\hat E}^{(+)}({\bf r}, t)|\Psi(a)\rangle \\
u_{det}({\bf r}, t)&= & a_0 \frac{\partial}{\partial a}  u_{mean}({\bf r}, t, a)  |_{a=0}
\end{eqnarray}
where $\mathcal{E}^{(1)}$ is the single photon electric field and $N$ the mean photon number. $a_0$ is the scaling factor necessary to normalize the mode $u_{det}$ to 1. $u_{det}$, called the \textit{detection mode}, characterizes the spatio-temporal distribution of the sensitivity of the optical system to a variation of the parameter. It can be used as the first mode of a new mode basis $\{{\bm u}_n\}$.

The determination of the Quantum Cram\'er-Rao bound $\Delta a_{QCR}$ in the case of a Gaussian state with high $N$ value is simplified, as one can show that the mean value of the field is $a$-dependent while its covariance noise matrix ${\bf \Gamma}_Q$ is $a$-independent   which leads to the following result \cite{Pinel2012a}
\begin{equation}\label{QCR}
\Delta a_{QCR}=\frac{a_0}{2 \sqrt{N}} \Delta_{det}
\end{equation}
where $\Delta_{det}$ is the quantum noise factor, equal to :
\begin{equation}
\Delta_{det}=\frac1{\sqrt{({\bf \Gamma}_Q^{-1})_{u_{det}}}}.
\end{equation}
$(\bf \Gamma_Q^{-1})_{u_{det}}$ being the first diagonal element  of the inverse covariance matrix in the detection mode. 

If the noise in this mode is not correlated with all the other modes of the basis, $\Delta_{det}$ is simply the r.m.s. value  of the quantum noise of a given quadrature of the detection mode: expression (\ref{QCR}) shows that \textit{the quality of the $a$ measurement is limited only by the quantum noise in the detection mode}. It is insensitive to the noises in all the other modes orthogonal to $u_{det}$ (in the general gaussian illumination case, one can show that the optimal use of ressources consists in populating the detection mode with the best squeezing source available, and in that case this mode is not correlated to the other modes  \cite{Pinel2012a}. $\Delta_{det}$ is equal to 1 when the $a$-encoding light quantum state is a coherent state. In this case  $\Delta a_{QCR}=a_0/2 \sqrt{N} $: this is the so-called standard Quantum Cramer Rao limit. $\Delta_{det}$ is below this value if one injects a squeezed vacuum state in the detection mode together with an intense coherent state in mode $u_{mean}$\cite{Fabre2000}. Expression (\ref{QCR}) also implies that injecting squeezed or EPR entangled states in other modes than $u_{det}$ will not decrease further the Quantum Cramer Rao bound: a single squeezed state is enough to reach the bound, provided it is put in the right mode, namely the detection mode \cite{Pinel2012a}. This implies also that it is not possible to accumulate the beneficial effects of squeezed states in two different modes for the measurement of a single parameter. Note that $\Delta a_{QCR}$ vanishes if one uses an infinitely squeezed vacuum state in the detection mode. Such a state is not physical, as it has an infinite energy. If one imposes a constraint of total finite energy $N\hbar \omega_0$,  equal to the sum of the mean energies of the modes $u_{mean}$ and  $u_{det}$, the limit scales as $N^{-3/4}$ \cite{Caves1981,Barnett2003}, an intermediate scaling between the standard quantum noise and the $N^{-1}$ Heisenberg scaling.

Actually, it is not enough to find the ultimate limits in parameter estimation. One needs of course to find a way to reach them.  Generally speaking, a  balanced homodyne detection with a Local oscillator put in the detection mode  allows to attain  the limit in all configurations \cite{Delaubert2008}, but there are in some cases more convenient ways, for example determine the $a$ estimator by computing combinations of photo-currents recorded in different parts of the illumination beam \cite{Treps2003}, or linear combinations of its spectral components, using the multiplexed detection outlined in section X E.  

Let us apply the approach that we just exposed to the well studied issue of a phase measurement using a Mach-Zehnder interferometer at mid-fringe. It involves two input modes $ f_1^{in}$ and $f_2^{in}$ incident on the first  beamsplitter, where we consider the usual case of $f_2^{in}$ being a vacuum mode, and two output modes $f_1^{out}$ and $ f_2^{out}$ at the output of the second beamsplitter. It is straightforward to show that in this configuration $u_{mean}= (f_1^{out} + f_2^{out})/\sqrt2$ and  $u_{det}= (f_1^{out} - f_2^{out})/\sqrt2$. These output modes respectively correspond, by back-propagation through the interferometer, precisely to the input modes $ f_1^{in}$ and $f_2^{in}$. It implies that the optimized configuration for an interferometric measurement using Gaussian states is to feed mode  $ f_1^{in}$ with an intense coherent state, and mode $ f_2^{in}$ with a squeezed vacuum state, a strategy that has been found years ago by C. Caves \cite{Caves1981} and currently implemented in gravitational wave antennas \cite{Schnabel2017}.  The approach outlined in this section therefore shows once again the optimized character of Caves' configuration and extends it to any optical measurement with multimode Gaussian states. It has been applied to the estimation of various parameters: transverse displacement and tilt of a $TEM_{00}$ light beam  \cite{Hsu2004,Delaubert2006} (for which the detection mode is the $TEM_{01}$ mode), time delay \cite{Lamine2008,Thiel2016}, propagation distance of a light pulse immune from atmospheric perturbations \cite{Jian2012}, mean frequency shift of broadband light, and transverse width of a highly focussed beam \cite{Chille2016}. Several experiments have shown that it is indeed possible to go beyond the standard Cram\'er-Rao bound by using squeezed light in the appropriate detection mode \cite{Treps2002, Pooser2015, Taylor:2013fo}. 

A general problem this approach can be naturally applied to is that of the ultimate resolution in optical imaging, a problem in which diffraction effects induce the existence of the so-called Rayleigh limit. It has been firstly tackled in the case of a coherent light image \cite{Kolobov2000}, and it has been shown that in this case injecting squeezing in appropriate modes (namely the prolate spheroidal ones) improves the optical resolution below the standard quantum noise \cite{Kolobov2008}, and that one could improve further the resolution by taking into account the sparsity of the image \cite{Wang2012}. It was more recently studied for incoherent illumination, more precisely to derive the ultimate resolution limit on the separation between two $TEM_{00}$ incoherent sources \cite{Tsang2016}.  It was shown that the Fisher information  contained in the intensity distribution in the image about this separation falls to zero as the separation drops below the Rayleigh limit, and that it is possible to extract more information on the separation  \cite{Tsang2016,Lupo2016}, and therefore to increase the accuracy of its estimation beyond the Rayleigh limit, by other measurement strategies \cite{Tsang2017}. This is a direct extension of the coherent estimation of a beam displacement to the incoherent case, and in particular one can make a homodyne measurement with LO in the $TEM_{01}$ Hermite Gauss mode  \cite{Yang2017}, project it on the detection mode which is the combination of two oppositely displaced $TEM_{00}$ modes \cite{Paur2016}, or demultiplex the amplitude image on the basis of spatial Hermite-gauss modes \cite{Tsang2017} by measuring the coefficients of the decomposition of the image field on such a mode basis. This can be achieved in particular using the MPLC device described in section XII . This scheme has been extended to the estimation of the axial separation between two point sources \cite{Zhou2019}, and to the time domain, using higher order Hermite-Gauss temporal modes to distinguish the arrival times of two incoherent ultrashort light pulses \cite{Donohue2018}. Furthermore, once multimode demultiplexing is available, one can use this information to perform multiparameter estimation, as was recently proposed in \cite{Napoli:2019do, Yu:2018jg}.

\section{Modes and states in quantum information processing}

\subsection{Measurement based quantum computing}

Multimode light is naturally at the heart of optical approaches to quantum computing, and while the objective of this review is not to address quantum information processing with light in general, we wish to emphasise the interest of using the tools introduced and used all along this review for quantum information processing techniques. In this context, even though the circuit based approach is still the most studied one, and for instance qubits can be successfully implemented on frequency-bins \cite{Lu2018}, extending Wavelength Division Multiplexing to the quantum domain, we will concentrate in this section on a recent paradigm for quantum computing, introduced in \cite{Raussendorf2001} as a {\it One-Way Quantum Computer} and more commonly named {\it Measurement Based Quantum Computing} (MBQC). The idea is to replace circuit based approach, where quantum gates are successively applied to the input qubits in order to perform a given operation, by a scheme where a large specific entangled state is generated and then successive measurements are performed on individual nodes of this state. More specifically, the input state should belong to a given class of graph states, the cluster states, and the result of each measurement is used to correct in a feedforward configuration the resulting state and to choose which observable will be measured in the next step. While first introduced in the qubit approach, MBQC was also extended to Continuous Variable domain \cite{Menicucci2006,Gu2009} with the advantage that cluster states in that regime can be deterministically generated. 

\subsection{Cluster states: concepts and experimental implementation}

We will focus here on the continuous variable approach, where, as we will see, the possibility to generate entanglement between modes through mode basis change allows for efficient and versatile generation of cluster states in which the nodes are precisely the different modes in multimode quantum light. These cluster states can be defined in an operational approach: applying on $N$ infinitely squeezed states a set of controlled-Z ($C_z$) gates defined by $C_z = e^{i\hat X_1 \hat X_2} $. This graph state structure is embedded in an adjacency matrix $V$, a $N\times N$ real symmetric matrix such that the multimode $C_Z$ gate writes:
\begin{equation}\label{czevolution}
C_z[V] = \prod_{1\leq j\leq k \leq N} \exp\left( iV_{jk} \hat X_j \otimes \hat X_k \right)
\end{equation}
In essence, the non-zero elements of $V$ induce a connection between two nodes of the graph. Usually, $V$ would be a matrix whose elements are either 0 or 1, but one can also have weighted graphs where non-zero elements can differ from 1.  

In order to study cluster state that can potentially be implemented experimentally, one first has to consider that the input states are finite squeezed states, the ideal cluster state being obtained in the limit of squeezing going to infinity. Then, because the evolution equation (\ref{czevolution}) corresponds to a quadratic Hamiltonian evolution applied to squeezed states, cluster states remain Gaussian states and can be constructed through symplectic transformations applied to the vacuum. As introduced in eq. (\ref{Q}), we name $\hat{\overrightarrow{Q}}_{sqz}$ the $2N$ elements column vector made of the quadrature operators for the initial squeezed states. One can show that the cluster state quadrature operators are given by \cite{Menicucci2006}:
\begin{eqnarray}
{\overrightarrow{\hat Q}}_{clu}\ = (C_z[V])^\dagger \, {\overrightarrow{\hat Q}}_{sqz}\, C_z[V]& = & \left( \begin{array}{cc}
\mathbb{1}_N & 0 \\
V & \mathbb{1}_N
\end{array}\right) {\overrightarrow{\hat Q}}_{sqz} \\
& = & \left( \begin{array}{cc}
\mathbb{1}_N & 0 \\
V & \mathbb{1}_N
\end{array}\right) K {\overrightarrow{\hat Q}}_{vac} \nonumber
\end{eqnarray}
Where $K=diag(\sigma_1, \sigma_2, \ldots, \sigma_N, \sigma^{-1}_1, \sigma^{-1}_2, \ldots, \sigma^{-1}_N)$ is the multimode squeezing matrix, with $\sigma_i>1$ (we consider input states squeezed on the $P$ quadrature). One finds that:
\begin{equation}
\lim_{\sigma_1, \sigma_2, \ldots, \sigma_N \rightarrow \infty}  \langle \left( {\overrightarrow{\hat P}}_{clu} - V {\overrightarrow{\hat X}}_{clu} \right)\rangle = 0
\end{equation}
The $N$ dimensional operator $ {\overrightarrow{\hat P}}_{clu} - V {\overrightarrow{\hat X}}_{clu} $ defines the $N$ nullifiers associated the graph, and hence governs its structure. It connects the $P$ quadrature of a node with the $X$ quadratures of the nodes connected to it. The nullifiers are quantities measurable experimentally. They are often used to assess the quality of a generated cluster state. In order to show that the experimentally generated state is indeed a cluster state, one needs to demonstrate that the nullifier variances are below the vacuum limit, but also that the graph shows multipartite entanglement as defined in section \ref{sec:multipartiteentanglement}.

The $C_z$ gate is hard to implement experimentally as its symplectic representation is not a simple basis change, but contains squeezing evolution. To bypass this difficulty, one can consider the total evolution from the vacuum state, which is symplectic, and calculate the corresponding Bloch Messiah decomposition (see section \ref{sec:BM}). Being a gaussian state, any approximated cluster state can be implemented with a multimode squeezer and a basis change $O_V$ acting on the vacuum. It was shown \cite{VanLoock2007} that the unitary matrix $U$ associated with the basis change $O_V$ as defined in (\ref{OrthogonalO}) is given by  the condition $\textrm{Re}(U) - V\textrm{Im}(U)=0$. Several experimental groups hence adopted the strategy to generate a set of independent squeezed states and implement the basis change $O_V$ corresponding to the desired cluster states \cite{Ukai2011,Su2012}, leading to cluster states with millions of modes\cite{Yokoyama2013}. One should note that the $O_V$ matrix is not unique, and in the case where the input squeezed states do not have the same squeezing level, the obtained nullifier variances depend on the actual $O_V$ which is used. Hence, optimising this matrix is important to reach the best cluster state for a given set of ressources \cite{Ferrini2013,Ferrini2016}. 

Being gaussian, and thus the result of a quadratic hamiltonian, cluster states can also be directly generated tailoring a non linear quadratic interaction, instead of being produced from independent parallel squeezers and basis change.  For instance, one can use the resonant frequency modes of an OPO as the nodes of the cluster, and tailor the entanglement by engineering the pump of the cavity. One can use several single frequency pumps \cite{Menicucci2008,Pfister2011,Chen2014} or a pulse shaped pump (\cite{Arzani2018}). It is also possible to cascade $\chi^{(3)}$ interaction in atomic vapours \cite{Jing2011,Pooser2014,Qin2014}.

\subsection{Non-Gaussian cluster states}

Cluster states are gaussian, and as such allow only for gaussian quantum information processing \cite{Weedbrook2012}, which cannot lead to quantum advantage~\cite{Bartlett:2002fo}, a property which can be extended to any positive Wigner function state~\cite{Mari:2012wi}. Thus an ingredient acting on the positivity of the Wigner function is required, which can take the form of a qubic gate~\cite{Gu2009}, or of a non-Gaussian encoding such as the so-called GKP encoding~\cite{Gottesman2001} which has been proven to allow for error corrected quantum computing protocols~\cite{Menicucci2014} with a requirement on squeezing now as low as 10$dB$~\cite{Fukui2018}. Several proposals for GKP state preparation have also been published \cite{Eaton:2019fp,Eaton2019,Weigand2018}.

In a spirit related to the scope of this review, another possibility is to generate non-gaussian cluster states that can then be used for universal quantum computing \cite{Sasaki2006, Quesada:2018cs, Gagatsos:2019es, Phillips:2019ho}. The most commonly used technique, in quantum optics, consists in adding or subtracting one, or several, photons to a Gaussian state \cite{Wenger:2004cw, Parigi:2007fva,Takahashi2008,OUrjoumtsev2007}. However, the challenge is to render this operation mode selective in a multimode context, a process that we will study in the following.

Let us consider the Wigner function $W_\Gamma(\overrightarrow{q})$ of a Gaussian state with covariance matrix ${\bf \Gamma}$, as defined in Eq.~(\ref{WignerGaussianMultimode}). This state can be a multimode entangled state for instance, that one wishes to {\it degaussify}, would it be for quantum information or quantum metrology purposes. We consider the coherent addition or subtraction of a single photon in a given mode ${\bm g}$, which amounts to the normalised application of the associated creation operator $\hat b^\dagger_g$ or  annihilation operator $\hat b_g$. This mode ${\bm g}$ can be any mode of the mode Hilbert space, and does not have to be one of the mode basis in which the Gaussian state is described. It can for instance be a non entangled mode, but also be a mode highly entangled to the rest of the system, such as the node of a cluster state. It can be shown that the Wigner function becomes~\cite{Walschaers2017}:
\begin{equation}\label{WignerDegaussifie}
W^{\pm}(\overrightarrow{q}) = \frac{1}{2}\left[ \overrightarrow{q}^{T*} {\bf \Gamma}^{-1}{\bf A}_g^{\pm}{\bf \Gamma}^{-1}\overrightarrow{q} -\mathrm{Tr}({\bf \Gamma}^{-1}{\bf A}_g^{\pm})+2 \right]W_\Gamma(\overrightarrow{q})
\end{equation}
where the operator ${\bf A}_g^{\pm}$ is the one amounting for the extra correlations induced by the single photon subtraction or addition operation, defined as:
\begin{equation}
{\bf A}_g^{\pm} = 2\frac{({\bf \Gamma}\pm \mathbb{1}){\bf P}_{g}({\bf \Gamma}\pm \mathbb{1})}{\mathrm{Tr}\left[({\bf \Gamma}\pm \mathbb{1}){\bf P}_{g}\right]}
\end{equation}
where ${\bf P}_{g}$ is the projector on the subspace associated with mode ${\bm g}$. For instance, when ${\bm g}$ is one of the modes of the mode basis, ${\bf P}_{g}$ reduces to a diagonal matrix with zero everywhere except for the two diagonal elements corresponding to the two quadratures associated with ${\bm g}$, where it is equal to 1. One should note that expression (\ref{WignerDegaussifie}) is valid for both pure and mixed states, and it can be easily applied to calculate the Wigner function of a subparty of a global state after photon addition or subtraction~\cite{Walschaers2017a}. In particular it can be used to study the negativity of the Wigner function, particularly simple in the case of non displaced states for which it can be probed in $\overrightarrow{q} = 0$ and amounts to determine the sign of the quantity $ 2 -\mathrm{Tr}({\bf \Gamma}^{-1}{\bf A}_g^{\pm}) $. 

As is well known, starting from a pure squeezed state, photon subtraction induces negativity, while whatever the input state photon addition induces negativity. The very same property does extend to graph states. However, the amount of negativity that is obtained depends more on the purity of the input state than on the quantity of squeezing available. In the multimode scenario, the complexity of this interplay can be examined using the above formula, in particular how non-Gaussianity spreads among a graph state, such as a cluster state. Remarkably, it can be shown for instance that if one removes a photon from a node of a cluster states, the non-gaussianity spreads up to two nodes away from the node on which the photon is subtracted~\cite{Walschaers2018}. This gives a method to induce fully non-gaussian cluster states for quantum information purposes. Finally, this formula can also be used to study how entanglement is induced by single photon operations. Starting from a pure state, one can simply evaluate the purity of the reduced state in a sub-space after the single photon operation. One can use this property to demonstrate that, for instance, the intrinsic separability of Gaussian states as defined in section \ref{intrinsic-separability} does not extend to photon added or photon subtracted Gaussian states. The structure of the state after single photon operation can be fundamentally different from the original Gaussian one, in the sense that for a well chosen mode of single photon operation the state becomes entangled whatever the mode basis.

In most experiments heralded single mode photon subtraction has been implemented using a weakly reflecting beamsplitter followed by a photon counter. When it detects a photon, one is sure that one photon has been removed from the transmitted beam. This process cannot be used for mode selective photon subtraction, because the beamsplitter and the photon detector do not discriminate between photons of different modes. As a result of this uncertainty the transmitted beam is in a mixed state. To get mode selectivity and significant negativity,  the authors of \cite{Ra2019} have used the process of sum frequency generation (SFG), which has already been presented in section XII in the context of mode conversion. They operated in the weak gate beam case, for which the evolution operator, given by (\ref{SFG}) is the same as the one for a weakly reflecting beamsplitter, but only for a single Hermite-Gauss input mode, the one which is identical to the Hermite-Gauss mode $HG_i$ of the gate beam. All the other Hermite-Gauss modes are not affected by the nonlinear process  \cite{Ra2017}. As a result, when one photon is recorded on the $ {\bm g}^{\prime}_{1}$ upconverted mode, one is sure that it has been removed from the $HG_i$ mode of the multimode input beam. The process is very flexible and can subtract a photon from linear superpositions of modes, from two-mode entangled states and from cluster states. When the device is fed at its input by the multimode quantum frequency comb described in section IX A4, significant levels of negativity are measured in these different configurations. The spreading of non-Gaussianity mentioned in the previous paragraph has also been observed.

Single mode photon addition can be implemented by parametric amplification in a $\chi^{(2)}$ nonlinear crystal: when one detects a photon on the idler mode, one is sure that one photon as been added to the input quantum state of the signal mode \cite{Zavatta2004}. The process has be extended to the photon addition to two temporal modes leading to micro-macro entanglement \cite{Biagi2018}, with a possibility of scalability to a larger number of modes.

\section{Conclusion}

The authors hope to have convinced the readers that considering in a comprehensive way the quantum states and the modes in which they are defined provides an interesting insight into many quantum optics issues and efficient ways to generate highly entangled quantum states. Increasing by a large factor the number of modes in an optical system does not pose intractable problems. Modes can be easily manipulated and computer controlled using Spatial Light Modulators, so that multimode quantum states, with well mastered mode shapes and correlations are promising scalable and reconfigurable carriers or processors of quantum information. Another advantage compared with the Discrete Variable approach is that highly entangled multimode states of light such as cluster states are generated in a deterministic, unconditional way whatever the size of the cluster state. But an important problem remains to be solved concerning the unconditional preparation of multimode non-Gaussian states.

\appendix

\section{Counting spatial modes in laser beams}

The number of spatial modes oscillating in a laser is an important parameter to characterize imperfect, non-single mode, laser beams \cite{Karny1983} . The number $M$, coming from the$"M^2 factor"$  introduced by Siegman (ref), is often considered as giving a direct measure of the number of transverse  modes. We would like to know whether this approach is compatible with the quantum one introduced in this paper.

 Let us call $k$ the wave vector in direction $x$. The $M^2$ factor is defined as:
\begin{equation}
M^2=\frac{\Delta x \Delta k}{(\Delta x \Delta k)_{min}}=2 \Delta x \Delta k
\end{equation}
$1/2$ being the minimum value of the product $\Delta x \Delta k$ allowed by the Fourier-Heisenberg inequality.

Let us now take as an example the case of a laser generating an incoherent superposition of $p$ Hermite-Gauss modes $h_n(x)$, with equal probabilities for each one and equal intensities in order to simplify the discussion. From a quantum mechanical point of view, it is described by the density matrix:
\begin{equation}
\rho= \sum_{n=1}^p \frac1p |\alpha : h_n \rangle \langle \alpha : h_n|
\end{equation}
where $ |\alpha  \rangle$ is a Glauber coherent state. The same reasoning as in section ??, based on the coherency matrix, tells us that $\rho$ describes indeed a quantum state having an intrinsic number of modes equal to $p$. We want now to know the relation between $p$ and $M$: as the coherency matrix is diagonal in the Hermite-Gauss mode basis, and using the properties of Hermite-Gauss modes, one has:
\begin{equation}
\Delta^2 x = \frac{\langle \int dx  x^2 {\hat N} \rangle}{ \langle \hat N \rangle}=\frac{ \sum_{n=0}^{p-1} \frac{| \alpha |^2}{p} (2 n + 1)w^2}{ \sum_{n=0}^{p-1}  \frac{|\alpha|^2}{p}}
\end{equation}
and
\begin{equation}
\Delta^2 k= \frac{\langle \int dx  k^2 {\hat N} \rangle}{ \langle \hat N \rangle}=\frac{ \sum_{n=0}^{p-1}  \frac{|\alpha|^2}{p}(2 n + 1)}{4 w^2 \sum_{n=0} ^{p-1} \frac{|\alpha|^2}{p} } 
\end{equation}
Knowing that $ \sum_{n=0}^{p-1} (2 n + 1)=p^2$, one finally finds:
\begin{equation}
\Delta^2 x= p^2 w^2 \quad ; \quad  \Delta^2 k= \frac{ p^2}{4 w^2}  \quad ; \quad  \Delta k \Delta x= \frac{ p^2}{2} 
\end{equation}
 so that $p$ is indeed equal to $M$ in this specific case, and close to this value in the general case.

  \section{Counting modes in parametric down converted light}
  
 The complexity of a bipartite quantum state, in particular its entanglement properties,  is related to the Schmidt number \cite{Guo2013,Dyakonov2014,Sharapova2015,Namiki2016}, i.e. the number of terms in the Schmidt decomposition. We would like to know whether the Schmidt number is indeed the mode number calculated from the coherency matrix.We will consider only a simple example, namely a bi-partite system the two parties A and B are multimode, for example the two-photon quantum state produced by parametric down conversion, written, in the 1D case, as: 
\begin{equation}
|\Psi \rangle =  \int dk_A dk_B g(k_A, k_B) | 1: k_A\rangle \otimes |1: k_B \rangle
\end{equation}
$g(k_A, k_B) $ containing the  phase matching and pump spatial properties, and $| 1: k_A\rangle$ being a single photon state of party A and wavevector $k_A$. Thanks to the Mercer theorem, $|\Psi \rangle$ can be written as a Schmidt sum:
\begin{equation}\label{Schmidt}
|\Psi \rangle =  \sum_{i=1}^S \lambda_i  | 1: u^i _A \rangle \otimes |1: u_B^i \rangle
\end{equation}
$u^i _A$ and $u^i _B$ being orthonormal eigenmodes  in parties A and B, $\lambda_i $ the corresponding positive eigenvalue, and $S$ the so-called Schmidt number. 

Let us call ${\hat b}^i_A$ and ${\hat b}^{i'}_B$ the annihilation operators in modes $u^i _A$ and $u^i _B$. We now use the  property that the mode number is the dimension of the space spanned by the vectors ${\hat b}^i_A|\Psi \rangle$ and ${\hat b}^{i'}_B|\Psi \rangle$. One has:
\begin{equation}
{\hat b}^i_A|\Psi \rangle = \lambda_i  |1: u_B^i \rangle \quad ; \quad {\hat b}^{i'}_B|\Psi \rangle = \lambda_{i'}  |1: u_A^{i'} \rangle
\end{equation}
All these vectors are orthogonal, meaning that the dimension of the generated space is $2S$. We have therefore shown that the number of modes is twice the Schmidt number.

\section{Modal dependence of the coincidence rate in the Hong-Ou-Mandel experiment}

We start from formula (\ref{Glauber2}), and chose two bases of modes, labeled  $\{ f^A_{m_A} \}$ for beam A and  $\{ f^B_{m_B} \}$ for beam B, in such a way the modes of same index ($m_A=m_B$)  are "mirror images"  of each other with respect of the beamsplitter. In this case, one can express simply the corresponding annihilation operators in function of the operators ${\hat b}^{A in}$ and  ${\hat b}^{B in}$ acting on the state before the beamsplitter, which is the factorized state  $| 1: g_A, 1: g_B \rangle$
\begin{equation}
 {\hat b}^A_{m}=\frac1{\sqrt2}({\hat b}^{A in}_{m}+e^{i \phi} {\hat b}^{B in}_{m}) \quad ; \quad {\hat b}B_{m}=\frac1{\sqrt2}({\hat b}^{A in}_{m} - {\hat b}^{B in}_{m}),
\end{equation}
the phase term $e^{i \phi}$ accounting for some delay, or some path diffference,  between the two arms of the interferometer. Using relation (\ref{annihil}) and the completeness relation for the mode basis, one gets the following expression for the normalized coincidence rate:
\begin{equation}
g^{(2)}=\frac12 \left( 1 -  |\overleftarrow{g}_A \cdot \overrightarrow{g}_B |^2 \cos  2 \phi \right)
\end{equation}
There are therefore no coincidences at zero path difference $\phi = 0$ when the modes of the two input photons are identical, spatially as well as temporally.

In the second configuration of the HOM interference, the input state generated by parametric down-conversion can be written, like in Appendix B, as the Schmidt sum (\ref{Schmidt}): 

\begin{equation}
|\Psi_2 \rangle = |0\rangle + \sum_{i=1}^S \sqrt{p_i}  | 1: g^i _A \rangle \otimes |1: g_B^i \rangle
\end{equation}
with $\sum_i p_i =1$. Using as previously mode bases that are mirror images of each other, one finds for the normalized coincidence rate at zero delay:
\begin{equation}
g^{(2)}(0)=\frac12 \left[ 1 - \sum_{i,j} \sqrt{p_i p_j} \,
(\overleftarrow{g}_A^i \cdot \overrightarrow{g}_B^j)( \overleftarrow{g}_A^j \cdot \overrightarrow{g}_B^i)^*\right]
\end{equation}
The coincidence rate vanishes when $ \overrightarrow{g}_A^{i} =\overrightarrow{g}_B^{i} \, \forall i$, i.e. when there is a \textit{perfect two-by-two matching for all the Schmidt modes of the  two beams}. This is achieved when there is total symmetry with respect to the exchange between the signal and idler A and B parts, i.e. when the matrix $G_{si}$ introduced in (\ref{Gsignalidler}) is symmetric.

One can now consider the same problem but with input coherent states of equal amplitudes and phase difference $\psi$,  $|\alpha \rangle$ and   $|\alpha e^{i\psi}\rangle$. A calculation analog to the previous one, based on formula (\ref{annihil2}), leads to the following result for $g^{(2)}(\phi=0)$:
\begin{equation}
g^{(2)}(0)=\frac12 \left( 1 -  |\overleftarrow{g}_A \cdot \overrightarrow{g}_B |^2 \cos 2 \psi \right)
\end{equation}
When the two modes are identical,  the "HOM dip"  is as expected zero for identical or opposite fields, and equal to 1/2 for incoherent fields. But one finds also a $100\%$ dip if $\psi$ takes randomly one of the two values $0$ and $\pi$, as it was recently stressed in \cite{Sadana2018}.

\begin{acknowledgments}

The authors would like to acknowledge in the first place the important contributions of all the PhD students and post-docs that worked successively in the quantum optics group of LKB. They would like to thank useful discussions with colleagues at the Laboratoire Kastler Brossel,  A. Bramati, E. Giacobino,  S. Gigan, J Laurat, C. Salomon among many others. Discussions with European colleagues in the different groups of European funded projects like QSTRUCT, QUANTIM, QCUMBER  have played a decisive role in the emergence of the concepts described in this review, in particular S. Barnett, A. Gatti, Y. Golubev, M. Kolobov, G. Leuchs, L. Lugiato, G.L. Oppo, G. de Valcarcel. We also acknowledge fruitful discussions at the occasion of conferences, visits and cooperation programs with  U. Andersen, H. Bachor, A. Furusawa, D. Horoshko, J. Howell, U. Leonhart, P. Lett,  M. Martinelli, P. Nussenzweig,  Z.Y. Ou, O. Pfister, Ping Koy Lam,  E. Polzik, M. Raymer, A. Sergienko, C. Silberhorn, W. Vogel, I. Walmsley and many others.

\end{acknowledgments}

\nocite{apsrev41Control}

\bibliography{RMP_Fabre_Treps}

\end{document}